\documentclass[AMA,STIX1COL]{WileyNJD-v2}

\articletype{Article Type}%

\received{26 April 2016}
\revised{6 June 2016}
\accepted{6 June 2016}

\usepackage{tcolorbox}
\usepackage{listings}
\newcommand{\code}[1]{\texttt{#1}}
\begin{document}

\title{On Code Reuse from StackOverflow: An Exploratory Study on Jupyter Notebook}

\author[1]{Mingke Yang}

\author[2]{Yuming Zhou}

\author[3]{Bixin Li}

\author[4]{Yutian Tang}

\authormark{YANG \textsc{et al}}

\address[1]{\orgdiv{School of Information Science and Technology}, \orgname{ShanghaiTech University}, \orgaddress{\state{Shanghai}, \country{China}}}

\address[2]{\orgdiv{Department of Computer Science and Technology}, \orgname{Nanjing University}, \orgaddress{\state{Nanjing}, \country{China}}}

\address[3]{\orgdiv{School of Computer Science and  Engineering}, \orgname{Southeast University}, \orgaddress{\state{Nanjing}, \country{China}}}

\address[4]{\orgdiv{School of Computer Science and Engineering}, \orgname{Nanjing University of Science and Technology}, \orgaddress{\state{Nanjing}, \country{China}}}

\corres{Yutian~Tang, Nanjing University of Science and Technology, China. \email{csytang@ieee.org}}

%\presentaddress{This is sample for present address text this is sample for present address text}

\abstract[Abstract]{Jupyter Notebook is a popular tool among data analysts and scientists for working with data. It provides a way to combine code, documentation, and visualizations in a single, interactive environment, facilitating code reuse. While code reuse can improve programming efficiency, it can also decrease readability, security, and overall performance. We conduct a large-scale exploratory study of code reuse practices in the Jupyter Notebook development community on the Stack Overflow platform to understand the potential negative impacts of code reuse. Our findings identified 1,097,470 Jupyter Notebook clone pairs that reuse Stack Overflow code snippets, and the average code snippet has 7.91 code quality violations. Through our research, we gain insight into the reasons behind Jupyter Notebook developers' decision to reuse code and the potential drawbacks of this practice.}


\jnlcitation{\cname{%
\author{Mingke Y},
\author{Yuming Z},
\author{Bixin L},
\author{Yutian T},
} (\cyear{2023}),
\ctitle{On Code Reuse from StackOverflow: An Exploratory Study on
Jupyter Notebook.}, \cjournal{Softw Pract Exper}.\cvol{2023}.}

\maketitle

\section{Introduction}

% The success of common open source software (e.g., Linux, LibreOffice) benefits from not only the efforts from developers, but also other publicly available software on the Internet. Code reuse has been widely adopted to be a key approach to delivering high quality software products in a timely manner \cite{wu2019developers}.

In the common paradigm of open source software development, the organization can be global and virtual, facilitated by the Internet and virtual communication among developers. The success of common open-source software, such as Linux and LibreOffice, relies not only on the efforts of developers but also on other publicly available software on the Internet. Code reuse is a widely adopted strategy for delivering high-quality software products efficiently. This approach allows developers to build upon existing work, rather than starting from scratch, to create new software solutions in a timely manner\cite{wu2019developers}. Code reuse can come from a variety of sources and in different forms, such as open source projects and software \cite{gharehyazie2017some}, Q\& A platform (e.g., Stack Overflow) \cite{abdalkareem2017code}. 

% \textcolor{red}{yuqi:tbd}
In recent years, the availability of large amounts of data and powerful computing resources led to the growth of fields such as data science and machine learning, where people in a wide range of fields, from healthcare and finance to social media and e-commerce are exposed to code \cite{kery2018story}. These people who are new to programming or data analysis tend to solve their coding problems through code reuse with the following purposes: (1) save development time by reusing code to implement specific functionality; (2) improve code efficiency by finding best practices; and (3) fix bugs in existing code by reusing bug-fixing code snippets. Therefore, it is common for inexperienced developers to focus on whether a code snippet meets their specific needs and implements the expected features, rather than on the overall quality of the code. This can be a problem as they may reuse poorly written or unreliable code, leading to errors, bugs, and other issues. 

\noindent\textbf{State-of-art.} Stack Overflow, the largest and most active online community for programmers and developers to ask and answer questions related to coding and software development, is a popular resource for code reuse. In the meantime, according to \cite{wang2020better}, Jupyter Notebook is the most widely used tool for analyzing data. Jupyter Notebook combines two components, a web application and notebook documents \cite{JupyterDoc}. The web application provides an interactive way to write notebook documents. Notebook documents contain codes, computational output, text and multimedia materials, it can serve as complete calculation records. As it provides a user-friendly interface and a powerful set of tools, it is particularly popular among inexperienced developers.Previous studies on Stack Overflow\cite{zhang2018code, ragkhitwetsagul2019toxic, fischer2017stack} and Jupyter Notebook\cite{wang2020better, koenzen2020code} were conducted in isolation. Specifically, Zhang et al.\cite{zhang2018code}. designed ExampleCheck to detect potential API usage violations in Stack Overflow posts, and Ragkhitwetsagul et al.\cite{ragkhitwetsagul2019toxic}. studied toxic code snippets in Stack Overflow. 
Fischer et al.\cite{fischer2017stack}. studied security issues caused by stack overflow snippets in android apps. Wang et al.\cite{wang2020better} experimentally demonstrated the existence of a large amount of poor-quality code in the Jupyter Notebook. Koenzen et al. \cite{koenzen2020code} investigated code reuse between Jupyter Notebook.

\noindent\textbf{Motivation.} However, to the best of our knowledge, there is no study on code reuse in Jupyter Notebook. In this paper, we conducted an exploratory study to fill this gap. Specifically, we provide evidence of the prevalence of code reuse in Jupyter Notebook, summarize why developers reuse code in Jupyter Notebook, and the impact of code reuse on code quality.

% 研究Jupyter Notebook中的Stack Overflow代码片段是十分重要的。Jupyter Notebook作为最广泛使用的分析数据的工具，它的用户包含大量缺少经验的开发者[]。当他们在编程上遇到困难，重用Stack Overflow上的代码片段时，往往更加在意代码片段是否实现了他们想要功能，而不是它的质量。这导致他们更有可能重用有问题的代码，从而对程序产生负面的影响。因此为了对Jupyter Notebook对Stack Overflow的代码重用有更深刻的见解,我们开展了一项探索性的研究。

% It is essential to study Stack Overflow code snippets for Jupyter Notebook. So to have a deeper insight into code reuse on Stack Overflow in Jupyter Notebooks, we conducted an exploratory study.

%Code reuse has a negative impact on the quality of the program \cite{koenzen2020code, huang2022towards}. However, code reuse in Jupyter Notebook is not well discussed. Therefore, research on code reuse in Jupyter Notebooks helps data analysis developers better understand how to reuse code. This paper presents an exploratory study of code reuse in Jupyter Notebook. Our research not only gives empirical evidence that code reuse exists in Jupyter Notebook but also provides some valuable insights into the potential reuse of Jupyter Notebook.

\noindent\textbf{Contribution.} In this paper, we make the following contributions: 

\noindent$\bullet $ First, to the best of our knowledge, we perform the \emph{first} large-scale exploratory study on Jupyter Notebook to explore the code reuse practices from the Stack Overflow platform for the development of Jupyter Notebook.

\noindent$\bullet $ Second, we conduct a systematic study on 3,758,196 Jupyter Notebook and 4,204,891 Stack Overflow code snippets.  We find 1,097,470 Jupyter Notebook clone pairs that reuse Stack Overflow code snippets. On average, a code snippet has 7.91 code quality violations.

\noindent$\bullet $ Third, we explore the reasons why Jupyter Notebook reuses Stack Overflow code snippets in terms of their Stack Overflow properties (e.g. whether they are accepted), developer experience, and developer purpose.

\noindent\textbf{Skeleton.} The rest of this paper is organized as follows: in Sec. \ref{sec:background}, we present the background and basic concepts in Jupyter Notebook and code reuse. In Sec. \ref{sec:methodology}, we present the research questions and describe the data collection process and the reuse detection methods. In Sec. \ref{sec:results}, we present the results and findings of our exploratory study. We discuss the lessons learned and threats to the validity of our work in Sec. \ref{sec:lessons}. In Sec. \ref{sec:relatedwork}, we introduce the related work of our study. In Sec. \ref{sec:conclusion}, we conclude our study.
\section{Background}\label{sec:background}
In this section, we briefly introduce several key concepts related to Jupyter Notebook, Stack Overflow, and code reuse. Furthermore, we leverage a running example to illustrate the code reuse practice with Stack Overflow.

\subsection{Jupyter Notebook}

A Jupyter Notebook contains a series of \textit{cells} \cite{Jupyter}. The type of a \textit{cell} can be a \textit{code} cell or a \textit{markdown} cell. A \textit{code} cell is a cell that contains executable source code. A \textit{markdown} cell is a rich text cell that supports markdown language, which is normally used to describe the code snippets in \textit{code} cells.

\begin{figure}[!htpb]
    \centering
    \includegraphics[width=0.5\textwidth]{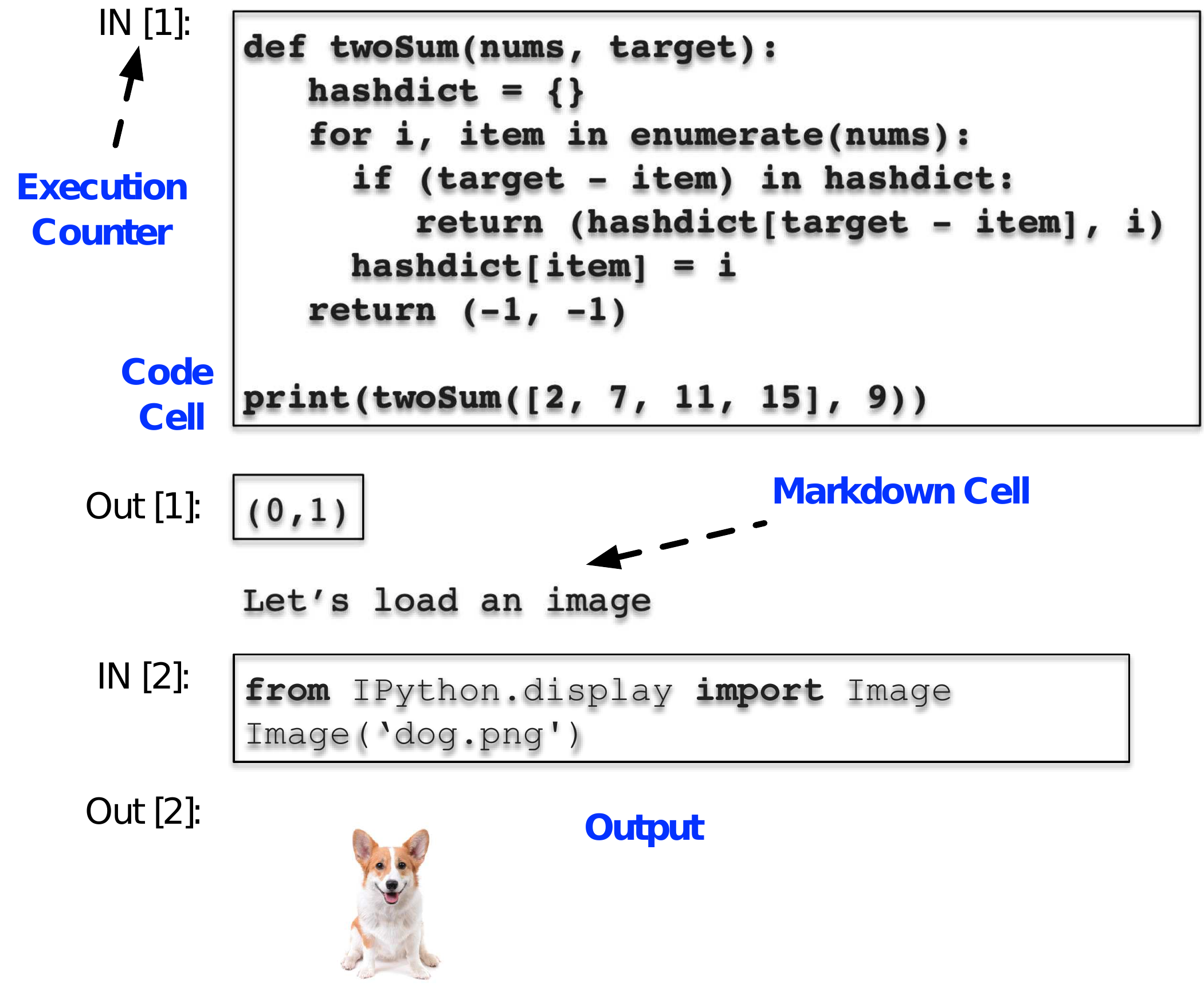}
    \caption{Jupyter Notebook Example}
    \label{fig:juypternotebook-example}
\end{figure}

Fig. \ref{fig:juypternotebook-example} illustrates a Jupyter Notebook example. It contains one markdown cell and two code cells. On the left of those cells, there are \textit{execution counters}, which show the execution order of these cells. The execution results of code cells are shown as outputs, which are displayed later. Note that the execution order only indicates the order of cells rather than any logical relationship between them.

\subsection{Code on Stack Overflow}

\begin{figure}[!htpb]
    \centering
    \includegraphics[width=0.5\textwidth]{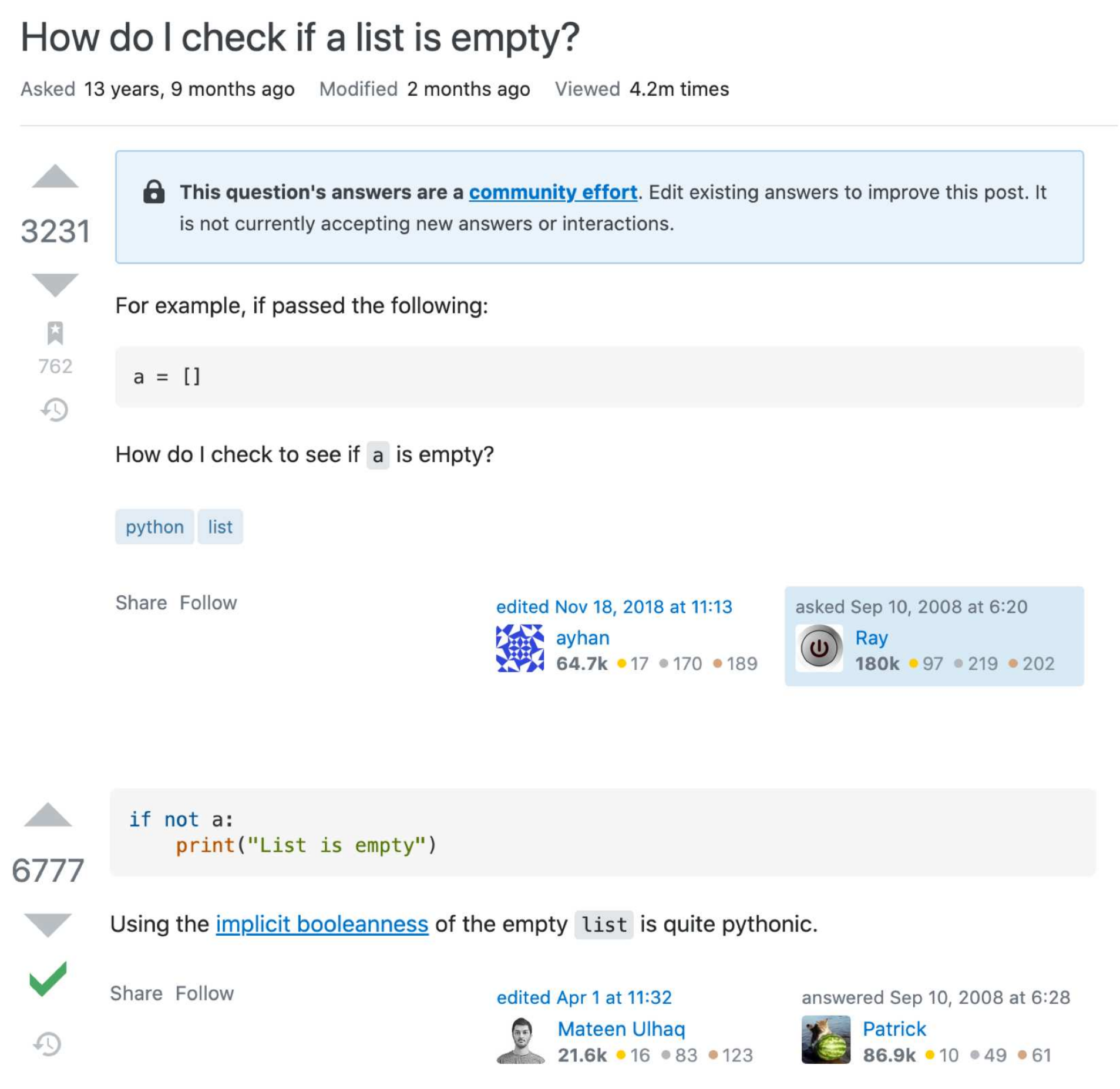}
    \caption{Stack Overflow Example}
    \label{fig:stackoverflow-example}
\end{figure}

Stack Overflow is one of the largest online Q\&A platforms for coding questions and answers \cite{StackOverflow}. In a Stack Overflow post, users can ask and answer questions and discuss the questions. Also, as shown in Fig. \ref{fig:stackoverflow-example}, askers can mark at most one answer as an ``accepted'' answer. Stack Overflow also allows developers to upvote and downvote answers. The number of votes for an answer is displayed as a score next to the answer.  For example, the score of the answer in Fig. \ref{fig:stackoverflow-example} is 6777, which indicates it is a high-quality solution to the question.

\subsection{A Motivating Example}
Existing research shows that developers leverage Stack Overflow code snippets to build their own programs \cite{meldrum2020understanding,ragkhitwetsagul2019toxic,wu2019developers}. In this work, we intend to study how developers reuse code from Stack Overflow to create Jupyter Notebook.

In what follows, we show the motivation of our research by describing a real-world example in which developers reuse code snippets from Stack Overflow in building Jupyter Notebook.

\begin{figure*}[!htpb]
    \centering
    \includegraphics[width=\textwidth]{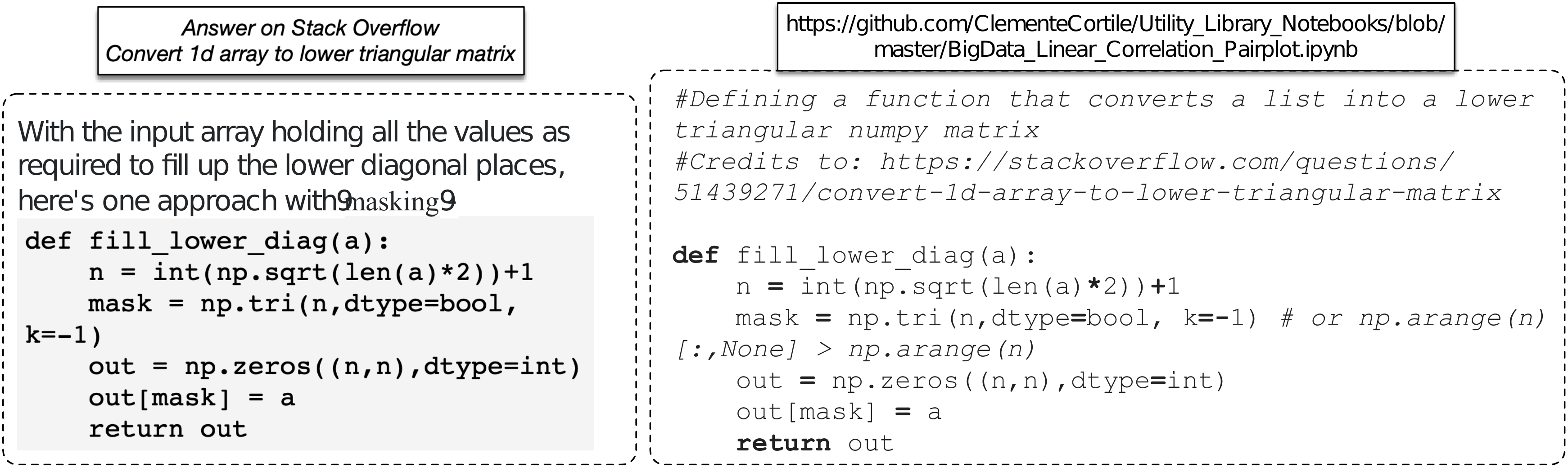}
    \caption{Source code from Stack Overflow is reused in building a Jupyter Notebook}
    \label{fig:motivating-example}
\end{figure*}

The left-hand side of Fig. \ref{fig:motivating-example} shows a code snippet (e.g., function \code{fill\_lower\_diag(a)}) from an answer on a Stack Overflow post. The answer shows how to convert a one-dimensional array into a lower, zero diagonal matrix while keeping all the digits. The right-hand side of Fig. \ref{fig:motivating-example} shows a code snippet from a real-world Jupyter Notebook named \code{BigData\_Linear\_Correlation\_Pairplot.ipynb}. 

By inspecting the timestamps of the commit and Stack Overflow, we can determine the code in the Jupyter Notebook reuses the code on the Stack Overflow post. In this Notebook, the function from Stack Overflow is directly reused by developers to convert a one-dimensional array into a lower, zero diagonal matrix. The general target of \code{BigData\_Linear\_Correlation\_Pairplot.ipynb} is to offer a series of functions to build a utility library to check the linear correlation between all variables in an extremely large dataset. Developers reuse the method \code{fill\_lower\_diag(a)} to convert the one-dimensional array into a matrix, which is later used to build a heat map.

With the aforementioned running example, we can conclude that identifying reused code snippets from Stack Overflow benefits the following aspects:

\noindent$\bullet$ First, we can measure how much code reuse there is in Jupyter Notebook;

\noindent$\bullet$ Second, code on Stack Overflow can contain potential defects, which can have negative impacts on other programs;

\noindent$\bullet$ Third, understanding code reuse practice assists us in determining the potential motivation for code reuse and improving code reuse practices.

\section{Research Questions \& Methodology} \label{sec:methodology}

\subsection{Research Questions}
% 参考问题 Toxic Code Snippets on Stack Overflow
% 
We perform an empirical study of online code clones
between Stack Overflow and Jupyter Notebook on GitHub to answer the following research questions (RQ):

\begin{itemize}
    % 相关文章 Stack Overflow: A Code Laundering Platform? RQ1，Toxic Code Snippets on Stack Overflow RQ2, Code Duplication and Reuse in Jupyter Notebook RQ1, Cross-project code clones in github RQ1, Stack Overflow in Github: Any Snippets There? RQ1
    
    \item \textbf{RQ1: How many code clone happened in Juypter Notebook?} To understand the code clone practice in Jupyter Notebook, we quantitatively measure the number of code clones between Stack Overflow and Jupyter Notebook on GitHub to understand the scale of the problem;
    
    \item \textbf{RQ2:Which part of the Stack Overflow post is used?} Intuitively, accepted answers or answers with high scores are more useful and trustworthy. In this RQ, we intend to investigate whether developers reuse non-accepted or low-scored code snippets. Investigating this RQ can assist Q\&A platforms in organizing answers;
    
    % On code reuse from StackOverflow: An exploratory study on Android apps
    \item \textbf{RQ3: Why do developers reuse code from Stack Overflow?} Developers are more prone to reuse code snippets from Stack Overflow. In this RQ, we intend to answer why developers reuse code and what the motivations are.

    % security
    % 相关文章：Understanding stack overflow code quality: A recommendation of caution RQ1-4
    \item \textbf{RQ4: What is the quality of code snippets provided in answers on Stack Overflow?} When reusing code snippets from Q\&A platforms like Stack Overflow, developers should consider the quality of the code snippets used. In this RQ, we intend to evaluate the quality of the code snippets from Stack Overflow. 
    
    % who reuses code 
    % 相关文章：Cross-project code clones in github RQ3, On code reuse from StackOverflow: An exploratory study on Android apps RQ3
    \item \textbf{RQ5: Who reuses code from Stack Overflow?} Previous research \cite{Ohad:2014} demonstrated that reusing code can have negative effects on building software. This always associates with less experienced developers. Here, we intend to examine who reuses Stack Overflow code among developers.
    
\end{itemize}

\subsection{Methodology}\label{subsec:methodology}

In this section, we present the pipeline for processing block-level code reuse between Stack Overflow and Jupyter Notebook projects on GitHub.

\subsubsection{Pipeline}
\begin{figure*}[!htpb]
    \centering
    \includegraphics[width=\textwidth]{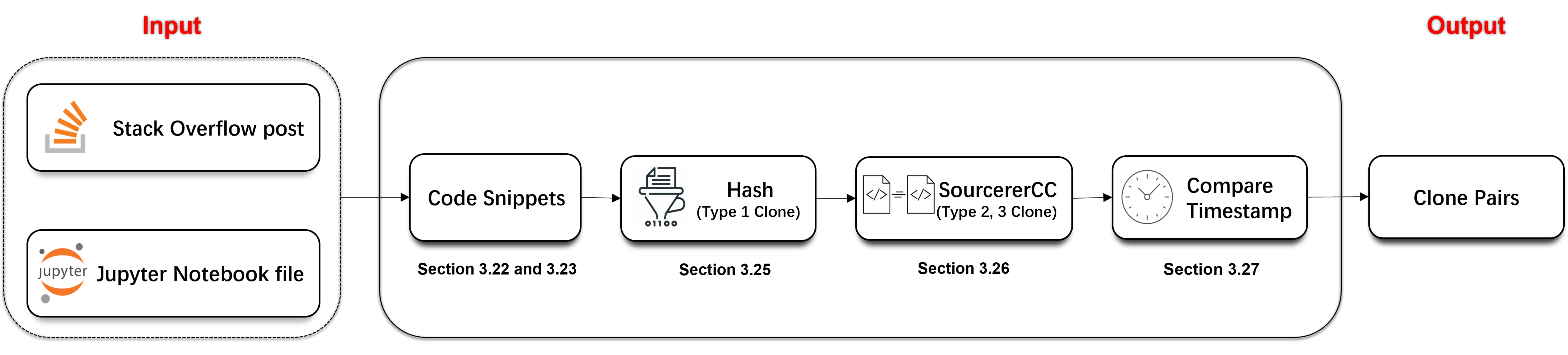}
    \caption{Pipeline of our study}
    \label{fig:pipeline}
\end{figure*}

We first briefly outline the pipeline of our approach. As shown in Fig. \ref{fig:pipeline}, we extract code snippets from Stack Overflow posts and Jupyter Notebook files. Then, we use MD5 to identify the Type-1 code clone pair and SourcererCC for the Type-2,3 code clone pair. Next, we use timestamps to get the clone pairs of Jupyter Notebook reused Stack Overflow code snippets. Finally, we analyze the clone pairs using different methods.

\subsubsection{Extract Code Blocks from Jupyter Notebook}\label{subsec:jupyternotebook-data}
To understand how code reuse happened in real-world Jupyter Notebook, we collect Jupyter Notebook from GitHub. Here, we follow the approach presented in \cite{pimentel2019large}. We collect Jupyter Notebook projects hosted on GitHub by specifying the identification language as ``Jupyter Notebook'' with the GitHub API. We collect Jupyter Notebook repositories built between 2013/01/01 and 2020/04/01. In summary, we obtain 188,302 repositories. For each Jupyter Notebook repository, we collect code snippets from each Notebook file (i.e., ends with \code{.ipynb}). The official Jupyter Notebook format is defined with the JSON schema. 

The following JSON scheme is used to render a code cell in a Jupyter Notebook. The \textit{cell\_type} attribute is \textit{code}, which represents the cell as a \textit{code} cell. The value \textit{source} attribute gives the source code in the cell. By parsing the Notebook files, we are able to extract the code snippets from the Notebook.

\begin{lstlisting}
{
  "cell_type" : "code",
  "execution_count": 1, # integer or null
  "metadata" : {
      "collapsed" : True, # whether the output of the cell is collapsed
      "scrolled": False, # any of true, false or "auto"
  },
  "source" : "[some multi-line code]",
  "outputs": [{
      # list of output dicts (described below)
      "output_type": "stream",
      ...
  }],
}
\end{lstlisting}

Finally, we obtain 3,758,196 code snippets.

\subsubsection{Extracting Code Blocks from Stack Overflow Posts}\label{subsec:stackoverflow-data}

We collect Stack Overflow posts from The SOTorrent dataset \cite{baltes2018sotorrent}.  The SOTorrent dataset collects the Q\&A posts from Stack Overflow. The latest version of SOTorrent contains the data updated to Dec,31,2020. We select Q\&A posts with the tag ``jupyter-notebook'' or ``python'' for our purpose. The tag ``jupyter-notebook'' gives all questions and answers related to Jupyter Notebook. As Python is the major programming language for Jupyter Notebook, we also collect posts with the tag ``Python''. In summary, we obtain 1,764,935 posts. The existing works \cite{abdalkareem2017code,nasehi2012makes} find that when the number of lines is greater than 5 it is more reasonable and meaningful. Thus, we adopt this setting to filter out code snippets within 5 lines. To be specific, for each post, we extract the code snippets from the post. The code snippets inside a post are embedded in the ``code'' tag. Then, we remove redundant code snippets by computing the MD5 of each code snippet. As a result, we obtain 4,204,891 code snippets.

\subsubsection{Type 1-4 Clones}
In general, code clones have been categorized into four types of clones: Type-1 to Type-4 \cite{roy2007survey}. Specifically, these four types of clones are:

\begin{itemize}
    \item \textbf{Type-1 Clone:} This type of clone refers to direct copy, and the only differences are spaces and comments. This type of clone can be detected easily;
    \item \textbf{Type-2 Clone:} Same as the Type-1 clone, Type-2 clone but also allows renaming variables;
    \item \textbf{Type-3 Clone:} Same as the Type-1 and Type-2 clone, Type-3 clone but also allows adding/deleting some statements;
    \item \textbf{Type-4 Clone:} Semantically identical, but not necessarily the same syntax.
\end{itemize}

In this paper, we focus on Type 1-3 clones. The following reason makes us discard Type-4 clone: in this paper, we aim to discuss how developers reuse code from Stack Overflow. For the Type-4 clone, it is hard to claim that developers reuse the code snippet from Stack Overflow only based on semantic similarity.

\subsubsection{Type-1 Clone}
To determine whether two code snippets are Type-1 clones, for two code snippets under comparison, we first remove all white spaces, including newlines and comments from the snippets. Second, we compute the MD5 hash for each code snippet. If two code snippets have the same MD5 hash, they are considered Type-1 clone pairs.

\noindent\textbf{Example.} Next, we leverage the following example to illustrate our Type-1 clone detection process. For the code snippets in List. \ref{lst:type1clone-a} and \ref{lst:type1clone-b}, we first remove all white spaces, including newlines and comments from snippets. Next, we compute the MD5 hash for the code in the List. \ref{lst:type1clone-a}. The result is ``00cee7ab3fa4839aad42c795415daa47''. We get the same MD5 hash for the code in the List. \ref{lst:type1clone-b}. Thus, we consider two code snippets are Type-1 cloned.

\begin{lstlisting}[caption={Type-1 Clone Example(a)},label={lst:type1clone-a}]
def max(a, b):
     if a > b:
         return a
     else:
         return b
\end{lstlisting}

\begin{lstlisting}[caption={Type-1 Clone Example(b)},label={lst:type1clone-b}]
# Return the maximum of two numbers
def max(a, b):
     if a > b:
        return a # a is the maximum
     else:
        return  b # b is the maximum
\end{lstlisting}

\subsubsection{Type-2 and Type-3 Clone}
To determine Type-2 and Type-3 code clone, we leverage SourcererCC \cite{sajnani2016sourcerercc} to detect such clones. SourcererCC is a token-based clone detection tool that supports Type-1 to Type-3 clones. As we already have presented how to detect Type-1 clones, here, we mainly use SourcererCC to detect Type-2 and Type-3 clones. 

SourcererCC performs clone detection with two phases: partial index creation and clone detection. In the first phase (partial index creation), SourcererCC parses the code blocks from the source files. Then, it tokenizes the code blocks with a simple scanner, which is aware of the token and block semantics of the target programming language (i.e., Python in our context). Next, it builds inverted index mapping tokens to blocks that contain them. It leverages a filtering heuristic to build a partial index of only a subset of the tokens in each block instead of building indexes for all tokens. Here, the filtering heuristic is used to reduce the candidates for comparison.

In the second phase (clone detection), SourcererCC iterates all code blocks. For each code block, it retrieves its candidate clone blocks from the index built in the previous phase. Furthermore, it leverages another filtering heuristic to compute the upper- and lower-bound similarity scores between the current block and its clone candidates. If the upper-bound similarity score for a candidate is lower than the predefined threshold, the process is terminated. If the lower-bound similarity score is higher than the predefined threshold, a cloned candidate is found. The aforementioned process is repeated until all cloned pairs are located.

We use the SourcererCC‘s default similarity threshold for clone detection, which is 80\%. That is, two code snippets are considered the clone pair if the similarity score is not less than 80\%.

\noindent\textbf{Example.} Next, we leverage a running example to illustrate the use of SourcererCC. 

\begin{lstlisting}[caption={Type-3 Clone Example},label={lst:type3clone}]
def max(a, b):
     if a > b:
        print()
        return a 
     else:
        return  b
\end{lstlisting}

Code snippets in List. \ref{lst:type1clone-a} and List. \ref{lst:type3clone} belong to the Type-3 clone, as the changes are added statements. The first step is tokenization. For each code snippet, the scanner parses the code snippet and outputs two files: \textit{files\_stats} and \textit{files\_tokens}. The \textit{files\_stats} file records statistics information of files, including file id, project id, project path, project URL, file hash, size bytes, lines, LOC (line of code), and SLOC (source line of code). The difference between LOC and SLOC in the SourererCC is that SLOC does not consider comment lines. The \textit{files\_tokens} file records lists of files together with various statistics and tokenized forms with the format: file id, project id, total tokens, unique tokens, token hash@ \#@token1@@::@@frequency, token2@@::@@frequency... Here, the ``token1@@::@@fre'' refers that the frequency of ``token1'' is ``fre''.  As a result, we obtain the following in the \textit{files\_stats} file:

\begin{lstlisting}[caption={files\_stats file Example}]
1,1,"1.zip/1/max.py","list1/max.py","94621.(hash value)..",73,5,5,5
2,2,"2.zip/3/max.py","list3/max.py","28fa9.(hash value)..",90,6,6,6
\end{lstlisting}

We obtain the followings in the \textit{files\_tokens} file:

\begin{lstlisting}[caption={files\_tokens file Example}]
1,1,12,7,e5d9...(hash value )..@\#@def@@::@@1,max@@::@@1,a@@::@@3,b@@::@@3,if@@::@@1,return...
2,2,13,8,70a6...(hash value )..@\#@def@@::@@1,max@@::@@1,a@@::@@3,b@@::@@3,if@@::@@1,...
\end{lstlisting}

Finally, SourcererCC outputs the indexes of cloned pairs.

\subsubsection{Compare Timestamp}
After obtaining the clone pair, we use the Stack Overflow API and Git command to acquire the timestamp of the code snippet's first appearance on Stack Overflow and GitHub. Then, we select clone pairs whose code snippets appeared on Stack Overflow earlier than Jupyter Notebook as target clone pairs. Specifically, for Stack Overflow, we use \url{https://api.stackexchange.com/2.3/posts/{POST_ID}/revisions?site=stackoverflow&filter=!6M2o(oKM-oyhS} to get edit revisions of the target Stack Overflow post. Then we use the clone code snippet to search its earliest appearance in the edit revision body and get the revision timestamp. For Jupyter Notebook, we first use \code{git log} to get all the commit ids and their corresponding commit dates. Then, we use \code{git checkout COMMIT\_ID} to update files in the specified commit. We search for the clone code snippet in the file where the clone is detected and get its earliest commit date.

\noindent\textbf{Example.} For the clone pair in Fig. \ref{fig:motivating-example}, we first query the Stack Overflow API, and the result is shown in List. \ref{lst:soquery}. By comparing the timestamps, we find that the revision with timestamp 1532079237 is the earliest, which is Jul 20, 2018. Then, we use \code{git log} to get the commits of the Jupyter Notebook project, the result is shown in List. \ref{lst:gitquery}. Next, we use the \code{git checkout COMMIT\_ID} to get the files in the specified commit. Searching for cloned code snippets, we find that Jul 13, 2019, is the matched commit date. By comparing the timestamps, we can determine the Jupyter Notebook project 
reused the code snippet on the Stack Overflow post.

\begin{lstlisting}[caption={Stack Overflow API query result},label={lst:soquery}]
{
    items: [
     {
         "creation_date":1532080688,
         "body": "...def fill_lower_diag(a):...",
     },
     {
          "creation_date": 1532079237,
          "body": "...def fill_lower_diag(a):...",
     }
    ]
...
}
\end{lstlisting}

\begin{lstlisting}[caption={Git command result},label={lst:gitquery}]
...
commit 55940698106310f8f5d8750060cf73b740c13e15
Date:   Sat Jul 13 14:34:54 2019 +0200
...
commit e8dd1d684b8122480d6bf100ee013dac156e17ec
Date:   Sat Jul 13 14:34:34 2019 +0200
...
\end{lstlisting}
\section{Research Questions and Discussion}\label{sec:results}

\subsection{RQ1: How many code clone happened in Juypter Notebook?}

% RQ1
% RQ3
% 
\begin{center}
\begin{table*}[b]%
\caption{Result of Clone Detection for Clone Pairs.\label{tab:clonepair}}
\centering
\begin{tabular*}{400pt}{@{\extracolsep\fill}lcc@{\extracolsep\fill}}
\toprule
&\multicolumn{2}{@{}c@{}}{\textbf{Clone Pairs}}  \\\cmidrule{2-3}
\textbf{Clone Type} & \textbf{Without Timestamp Comparison}  & \textbf{With Timestamp Comparison}     \\
\midrule
Type-1 & 71,851  & 33,448    \\
Type-2,3 & 1,923,943  & 1,064,022     \\
All Clone Type & 1,995,794  & 1,097,470      \\
\bottomrule
\end{tabular*}
\end{table*}
\end{center}

\begin{center}
\begin{table*}[b]%
\caption{Result of Jupyter Code Snippets in Clone pairs.\label{tab:clonedetectionjupyter}}
\centering
\begin{tabular*}{400pt}{@{\extracolsep\fill}lcc@{\extracolsep\fill}}
\toprule
&\multicolumn{2}{@{}c@{}}{\textbf{Jupyter Code snippets}}  \\\cmidrule{2-3}
\textbf{Clone Type} & \textbf{Without Timestamp Comparison}  & \textbf{With Timestamp Comparison}     \\
\midrule
Type-1    & 30,138   & 19,638  \\
Type-2,3  & 272,560   & 184,727  \\
All Cloned Snippets       & 285,545    & 193,248\\
\bottomrule
\end{tabular*}
\end{table*}
\end{center}

\noindent\textbf{Motivation.} To understand the code clone practice in Jupyter Notebook, in this RQ, we quantitatively measure the number of code clones between Stack Overflow and Jupyter Notebook on GitHub to understand the scale of the problem.

\noindent\textbf{Methodology.} The methodology for this RQ is presented in Sec. \ref{subsec:methodology}.

\noindent\textbf{Results.} The result of our clone pair detection is shown in the Table. \ref{tab:clonepair}. We find 1,995,794 clone pairs containing 83,747 stack overflow code snippets and 285,545 Jupyter Notebook code snippets. The number of code snippets is less than the number of clone pairs because there is a situation where a code snippet is reused multiple times. These clone pairs also relate to 71,627 Stack Overflow posts and 59,942 Github repositories, representing 4.06\% of collected posts and 31.83\% of collected repositories. We summarize the number of Jupyter Notebook code snippets in cloned pairs, shown in the Table. \ref{tab:clonedetectionjupyter}.

For \textbf{Type-1 Code Clone}, we find 58,446 Type-1 clone pairs. We use the hash value to filter out identical code snippets and then count the number of distinct code snippets. Interestingly, there are only 5,472 distinct Stack Overflow code snippets, much smaller than the total number of clone pairs we have found. On average, every code snippet is reused 10.68 times. Using the timestamp of the code snippet first shown in the Jupyter Notebook file and Stack Overflow, we find 33,448 code clone pairs are Jupyter Notebook code snippets that reused code from Stack Overflow posts. 

Next, we present the top five code snippets commonly reused with Type-1 code clone. To focus on code that implements specific function. We ignore meaningless code snippets such as importing libraries and defining initial data.  The most commonly reused code snippet is for setting up Jupyter Notebook to hide input code and only show outputs and markdown, shown in List. \ref{lst:mostreusedcode}. This code snippet was reused 273 times. When a Jupyter Notebook developer shares the notebook with others but only wants to share the results and text rather than the code. This code is reused and placed at the beginning of the notebook.
\begin{lstlisting}[caption={Most Type-1 reused code},label={lst:mostreusedcode}]
from IPython.display import HTML

HTML('''<script>
code_show=true; 
function code_toggle() {
 if (code_show){
 $('div.input').hide();
 } else {
 $('div.input').show();
 }
 code_show = !code_show
} 
$( document ).ready(code_toggle);
</script>
<form action="javascript:code_toggle()">
<input type="submit" 
value="Click here to toggle on/off the raw code."></form>''')
\end{lstlisting}

As shown in List.\ref{lst:secondemostreusedcode}, the second most reused code(243 times) loads detection graph from the external directory. When training a TensorFlow model, it is common to save it as a .pb (protocol buffers) file. Protocol Buffers is a method of serializing data\cite{protocol-buffers}, in this case for saving graph definitions and model weights. This code is reused when developers want to use the model defined in the protobuf file.

\begin{lstlisting}[caption={Second most Type-1 reused code},label={lst:secondemostreusedcode}]
detection_graph = tf.Graph()
with detection_graph.as_default():
  od_graph_def = tf.GraphDef()
  with tf.gfile.GFile(PATH_TO_CKPT, 'rb') as fid:
    serialized_graph = fid.read()
    od_graph_def.ParseFromString(serialized_graph)
    tf.import_graph_def(od_graph_def, name='')
\end{lstlisting}

The third most commonly reused code snippet(227 times) is for adding custom CSS style for notebook, is shown in List. \ref{lst:thirdemostreusedcode}. This code is reused when the developer is not satisfied with the default style of Jupyter Notebook. This code is often placed at the end of the notebook to isolate it from the data analyze code.

\begin{lstlisting}[caption={Third most Type-1 reused code},label={lst:thirdemostreusedcode}]
#Apply styles
from IPython.core.display import HTML

def css_styling():
    styles = open("styles/custom.css", "r").read()
    return HTML(styles)
css_styling()
\end{lstlisting}

The fourth most commonly reused code snippet is to get the files uploaded to Colab and print it, as shown in List. \ref{lst:forthemostreusedcode}. This code snippet reused 227 times. Google Colab is a free Jupyter notebook environment that runs in the cloud. The code snippet is used to query which file has been uploaded (usually data files), and will be used later.

\begin{lstlisting}[caption={Forth most Type-1 reused code},label={lst:forthemostreusedcode}]
from google.colab import files
uploaded = files.upload()
for fn in uploaded.keys():
  print('User uploaded file "{name}" with length {length} bytes'.format(
      name=fn, length=len(uploaded[fn])))
\end{lstlisting}

Last, the fifth most reused code snippet(222 times) is shown in the List. \ref{lst:fifthmostreusedcode}. This code snippet is used \code{StandardScale} to scale the dataset to unit variance. After loading, the data may need to be preprocessed using normalization. This code snippet is reused when data needs to be scaled to the standard normal distribution.

\begin{lstlisting}[caption={fifth most Type-1 reused code},label={lst:fifthmostreusedcode}]
from sklearn.preprocessing import StandardScaler
sc = StandardScaler()
X_train = sc.fit_transform(X_train)
X_test = sc.transform(X_test)
\end{lstlisting}

After introducing the 5 most common Type-1 reuse code snippets, we find that the reason for reuse can be divided into two categories:

\begin{itemize}
    \item Customise Jupyter Notebook:
    List.\ref{lst:mostreusedcode} sets the visibility of code cells. List.\ref{lst:thirdemostreusedcode} using external CSS to custom Jupyter Notebook theme. As the functionality of these code snippets is not related to data analysis, unskilled developers may lack knowledge about this part, thus they reuse code snippets from Stack Overflow.
    
    \item Prepare for data analysis:
    List.\ref{lst:secondemostreusedcode} loads the trained model. List.\ref{lst:forthemostreusedcode} queries the files that have been uploaded to the cloud. List.\ref{lst:fifthmostreusedcode} normalizes the data for preprocessing. They are off-the-shelf implementations of these functions on Stack Overflow, so developers reuse these code snippets.
\end{itemize}

For \textbf{Type-2,3}, compared to Type-1 code clone, the detection criteria of Type-2,3 code clone are less stringent. We find 1,937,348 pairs of Type-2,3 clones. This is about 33.15 times the number of Type-1 clone pairs and corresponds to 97.07\% of the clone pairs. Similarly, we find 1,064,022 clone pairs are Stack Overflow posts code snippets reused by Jupyter Notebook from the Type-2,3 Code Clone pairs. These clone pairs consist of 65,811 Stack Overflow code snippets and 184,727 Jupyter Notebook code snippets, representing 1.60\% of the Stack Overflow code snippets and 4.91\% of the Jupyter Notebook code snippets we collected. 
It is remarkable that 1,097,470 clone pairs are the Jupyter Notebook reuse Stack Overflow posts.

Similarly, we present the top five code snippets commonly reused with Type-2,3 code clone. The most common code reuse snippets for Type 2,3  is shown in List.\ref{lst:mostreusedcode23}. This code snippet reused 3601 times. It is used to set up the notebook so that figures are displayed inline in the notebook, and initialize the plotting settings. This code is reused when developers need to set up \code{matplotlib}. In Type 2,3 code reuse, developers usually change the \code{pyplot}'s runtime configuration options(rcParams).

\begin{lstlisting}[caption={Most Type-2,3 reused code snippets},label={lst:mostreusedcode23}]
import random
import numpy as np
from cs231n.data_utils import load_CIFAR10
import matplotlib.pyplot as plt
%matplotlib inline 
plt.rcParams['figure.figsize'] = (10.0, 8.0) # set default size of plots
plt.rcParams['image.interpolation'] = 'nearest'
plt.rcParams['image.cmap'] = 'gray'
\end{lstlisting}

In supervised machine learning, to prevent overfitting, the dataset is usually divided into two parts: the training set and the test set. The second most frequently reused code (1845 times) snippet is to load such dataset and print the shape of the dataset, as shown in the List. \ref{lst:secondmostreusedcode23}. Supervised learning is the machine learning task of inferring functions from a labeled dataset. When developers need to import the dataset, this code is reused. The sanity check is performed by printing the size of the dataset. A common change to use this code snippet is to modify the dataset path or dataset name.

\begin{lstlisting}[caption={Second most Type-2,3 reused code snippets},label={lst:secondmostreusedcode23}]
cifar10_dir = 'cs231n/datasets/cifar-10-batches-py'
X_train, y_train, X_test, y_test = load_CIFAR10(cifar10_dir)

print 'Training data shape: ', X_train.shape
print 'Training labels shape: ', y_train.shape
print 'Test data shape: ', X_test.shape
print 'Test labels shape: ', y_test.shape
\end{lstlisting}

After importing data, a common next step is to reshape the data into a form that facilitates the rest of the data analysis. As shown in List.\ref{lst:thirdmostreusedcode23}, the third most reused code (1285 times) is formatting the dataset shape and printing size of the dataset. Similar to the List \ref{lst:secondmostreusedcode23}, this code snippet performs reshape on multiple datasets, training the model with supervised learning. This code snippet is used to pre-process the data. Then the dataset size is printed for sanity check. A common way to reuse Type 2,3 for this code snippet is to change the parameters in \code{reshape}.

\begin{lstlisting}[caption={Third Most Type-2,3 reused code snippets},label={lst:thirdmostreusedcode23}]
def reformat(dataset, labels):
  dataset = dataset.reshape((-1, image_size * image_size)).astype(np.float32)
  labels = (np.arange(num_labels) == labels[:,None]).astype(np.float32)
  return dataset, labels
train_dataset, train_labels = reformat(train_dataset, train_labels)
valid_dataset, valid_labels = reformat(valid_dataset, valid_labels)
test_dataset, test_labels = reformat(test_dataset, test_labels)
print('Training set', train_dataset.shape, train_labels.shape)
print('Validation set', valid_dataset.shape, valid_labels.shape)
print('Test set', test_dataset.shape, test_labels.shape)
\end{lstlisting}

The fourth most commonly reused code snippet(1150 times) is for plotting the confusion matrix, as shown in List. \ref{lst:fourthemostreusedcode23}. This code snippet is reused when users need to draw the confusion matrix. The common reused method of Type-2,3 is to change the plot drawing options without changing the data processing part.

\begin{lstlisting}[caption={Fifth Most Type-2,3 reused code snippets},label={lst:fourthemostreusedcode23}]
def plot_confusion_matrix(cm, classes,
                          normalize=False,
                          title='Confusion matrix',
                          cmap=plt.cm.Blues):
    plt.imshow(cm, interpolation='nearest', cmap=cmap)
    plt.title(title)
    plt.colorbar()
    tick_marks = np.arange(len(classes))
    plt.xticks(tick_marks, classes, rotation=45)
    plt.yticks(tick_marks, classes)

    if normalize:
        cm = cm.astype('float') / cm.sum(axis=1)[:, np.newaxis]
        print("Normalized confusion matrix")
    else:
        print('Confusion matrix, without normalization')

    print(cm)

    thresh = cm.max() / 2.
    for i, j in itertools.product(range(cm.shape[0]), range(cm.shape[1])):
        plt.text(j, i, round(cm[i, j],4)*100,
                 horizontalalignment="center",
                 color="white" if cm[i, j] > thresh else "black")

    plt.tight_layout()
    plt.ylabel('True label')
    plt.xlabel('Predicted label')
\end{lstlisting}

Last, the fifth most commonly reused code snippet (1008 times) is shown in the List. \ref{lst:fifthmostreusedcode23}. This code snippet is the same as List. \ref{lst:thirdemostreusedcode}, change the style of the notebook by reusing this code snippet. Changing the parameters in the \code{open} function is a common way to reuse it in Type-2,3 Clone.

\begin{lstlisting}[caption={Fifth most Type-2,3 reused code snippets},label={lst:fifthmostreusedcode23}]
from IPython.core.display import HTML
def css_styling():
    styles = open("./example.css", "r").read()
    return HTML(styles)
css_styling()
\end{lstlisting}

After introducing the 5 most common Type-2,3 reuse code snippets, we find that the reason for reuse can be divided into three categories:

\begin{itemize}
    \item Customise Jupyter Notebook: List. \ref{lst:fifthmostreusedcode23} loads external CSS to modify Jupyter Notebook's style settings. The reason for reusing this code snippet is the same as Type-1 code reuse.
    
    \item Prepare for data analysis: List. \ref{lst:secondmostreusedcode23} loads the dataset and prints the dataset size. List. \ref{lst:thirdmostreusedcode23} reshapes the dataset. The reason for reusing those code snippets is the same as Type-1 code reuse.
    
    \item Plot setting: List. \ref{lst:mostreusedcode23} sets the \code{pyplot} runtime configuration options. List. \ref{lst:fourthemostreusedcode23} plots the confusion matrix by setting \code{pyplot}. Plotting the data is an important part of data analysis. However, there are some common parameters that need to be set when plotting, such as \code{figsize} or \code{title}. Developers look for help from Stack Overflow to set those parameters. Since different developers have different ways of presenting data, they may have different settings for the plot. Therefore this part of the function is often used with Type-2, 3 code reuse instead of Type-1 code reuse.
\end{itemize}

\begin{tcolorbox}

\section*{Answer to RQ1 (Code Reuse Practice)}
By investigating 3,758,196 Jupyter Notebook and 4,204,891 Stack Overflow code snippets, we find 1,995,794 clone pairs containing 83,747 stack overflow code snippets and 285,545 Jupyter Notebook code snippets. The ratio of Type-1  to Type-2,3 code clones is about 1:33.15. Among all clone pairs, 33,448 Type-1 and 1,064,022 Type-2,3 clone code pair is the code snippets from the Stack Overflow post and reused by Jupyter Notebook code snippets. We find that customizing Jupyter Notebook and preparing for data analysis are the reasons for most Type-1 reusing code snippets. The reason most reused code snippets are reused by Type-2,3 is the same as Type-1 code reuse but with adding settings for the plot.
\end{tcolorbox}

\subsection{RQ2: Which part of the Stack Overflow post is used?}

% On code reuse from StackOverflow: An exploratory study on Android apps

\noindent\textbf{Motivation.} Intuitively, we suppose that accepted solutions or answered solutions with high votes are more useful and trustworthy. In this RQ, we intend to investigate whether developers use non-accepted answers or answers with low scores. Meanwhile, we also intend to investigate whether exists relationships between the types of clones and the quality of answers. For example, a high-quality solution may be reused without any modification (i.e., Type-1 clone).

\noindent\textbf{Methodology.} To cope with this, in this RQ, we check which parts (e.g., accepted answer, non-accepted answer, questions) of the Stack Overflow post are reused by developers. We check whether the reused code snippet is with the highest score (a.k.a. votes). Note that vote can be a positive value or a negative value. A negative value indicates the number of negative votes is larger than the number of positive votes. While a positive value indicates, the number of positive votes is larger than the number of negative votes. Finally, we ask two authors of this paper to categorize the motivations for reusing those parts of code snippets. If there exist discrepancies, another author of this paper is involved in the discussion until a consensus is reached.

\noindent\textbf{Results.}
We count the number of code snippets of reused code snippets in different post sections. The result is shown in Table. \ref{tab:codecomefrom}. 

\begin{center}
\begin{table*}[t]%
\caption{Number of reused code snippets from different sections in Stack Overflow.\label{tab:codecomefrom}}
\centering
\begin{tabular*}{500pt}{@{\extracolsep\fill}lccD{.}{.}{3}c@{\extracolsep\fill}}
\toprule
&\multicolumn{2}{@{}c@{}}{\textbf{Frequency}} & \multicolumn{2}{@{}c@{}}{\textbf{Average Vote}} \\\cmidrule{2-3}\cmidrule{4-5}
\textbf{Source} & \textbf{Type-1}  & \textbf{Type-2,3}  & \multicolumn{1}{@{}l@{}}{\textbf{Type-1}}  & \textbf{Type-2,3}   \\
\midrule
Question & 2,485 (3.55\%)  & 37,339 (53.28\%)  & 17.89  & 12.53   \\
Accepted answer & 631 (0.90\%)  & 10,756 (15.35\%)  & 35.56  & 12.03   \\
Non-accepted answer & 1,148 (1.64\%)  & 17,716 (25.28\%)  & 11.25  & 5.33   \\
\bottomrule
\end{tabular*}
\end{table*}
\end{center}

\noindent$\bullet$\textbf{Reuse question section code snippets} We observe 2485 (3.55\%) code snippets reused by Type-1 from the question section and 37,339  (53.28\%) for Type-2,3 code reuse. There are more reused code snippets from the question section than the answer section.

By manually checking for reused code snippets in Question section. we summarize the reasons that developers reuse them.

\begin{itemize}
    \item Demo/Tutorial code snippets: These code snippets come from other sources, such as sample code from official documentation, libraries, or books. From the questioner's point of view, they often ask questions about how it works. From the developers' point of view, the code snippet meets their needs;
    \item  Reusing test cases: This code snippet is test code written by the questioner, and other developers also reuse this code snippet to test specific features;
    \item  Fixing buggy code: The questioner demonstrated a buggy code snippet in the question section. The developers reuse it by fix the buggy codes. For example, as shown in List. \ref{lst:bugquestioncode}, a buggy code is provided in the question section, due to Python's If statement selects the branch which first satisfies the condition. So when the input is 120, the output is \code{HOT} instead of \code{REALLY HOT!}. The bug in the code snippet has been fixed in Jupyter Notebook by reordering the If branches, shown in the List. \ref{lst:correctnotebookcode}.
    
\begin{lstlisting}[caption={Buggy code snippet in question section},label={lst:bugquestioncode}]
temp = 120
if temp > 85:
   print("Hot")
elif temp > 100:
   print("REALLY HOT!")
elif temp > 60:
   print("Comfortable") 
else:
   print("Cold")
\end{lstlisting}
\begin{lstlisting}[caption={Reused code snippet in Jupyter Notebook},label={lst:correctnotebookcode}]
temp = 120
if temp > 100:
   print "REALLY HOT!"
elif temp > 85:
   print "Hot"
elif temp > 60:
   print "Comfortable" 
else:
   print "Cold"
\end{lstlisting}
\end{itemize}

The average votes for code snippets from the Question section are 17.89 for Type-1 and 12.53 for Type-2,3. However, the average vote for the entire question section we collected is 9.28.

\noindent$\bullet$\textbf{Reuse accepted answer code snippets} For reused code snippets in the accepted answer question, we find 631 (0.90\%) Type-1 code reuse and 10,756 (15.35\%) Type 2,3 code reuse. Code snippets from the accepted answers represent 37.84\% of all answers. The average acceptance rate of all the answers in our data is 34.57\%, which shows that developers do not always reuse code from accepted answers. 

We summarize the reasons that developers reuse the accept answer section code snippet.

\begin{itemize}
    \item Provide developer requirement features: These code snippets implement the functionality mentioned in the question;

    \item Fix the question section's code bug: The accepted answer fixes the buggy code in the question section. The developer reuses this code to fix the same bug; For example, as shown in List. \ref{lst:bugcode}, The questioner uses \code{str()} to convert from Unicode to UTF-8 text, which throws an UnicodeEncodeError. The accepted answer shows that the questioner should use \code{.encode()} to encode the string, as shown in List. \ref{lst:correctcode}.

\begin{lstlisting}[caption={Buggy code in question section},label={lst:bugcode}]
p.agent_info = str(agent_contact + ' ' + agent_telno).strip()
\end{lstlisting}

\begin{lstlisting}[caption={Correct code snippet in accepted answer section},label={lst:correctcode}]
p.agent_info = u' '.join((agent_contact, agent_telno))\
                .encode('utf-8').strip()
\end{lstlisting}

    \item API Usage: It demonstrates how to use the library's APIs. Developers reuse these code snippets to invoke the same API;
\end{itemize}

We collect an average of votes for all code snippets from answer sections 2.97. Answers containing reused code have an average of votes higher than that, regardless of the reuse Type of code snippet.

\noindent$\bullet$\textbf{Reuse non-accepted answer code snippets} For code snippets in non-accepted answer, there are 1,148 (1.64\%) reused by Type-1 and  17,716 (25.28\%) reused by Type 2,3.

The reasons developers reuse code snippets in the non-accepted answer section are summarized below.

\begin{itemize}
    \item Different code version: The version of the accepted answer's code does not match the developer's needs. For example, the code snippet in the accept answer is written in Python 2. Developers have to reuse the code segment in Python 3 from a non-accept answer;
    
    \item More generic code: The accepted answer to this question fit the questioner's requirements. However, this answer is specific to the questioner's problem. It does not easily transfer to other problems, but the non-accepted code snippet gives a generic way to solve the problem. For example, in the question section, the questioner asked how to concatenate \code{listone = [1, 2, 3]}, \code{listtwo = [4, 5, 6]} the two lists in the python. The accepted answer shown List.\ref{lst:twolists} use \code{+} operator to solve this problem. However, the non-accept answer provides a function to combine multiple lists, shown in the List. \ref{lst:multiplelists}.
    
\begin{lstlisting}[caption={Combine two lists},label={lst:twolists}]
listone = [1,2,3]
listtwo = [4,5,6]

listthree = listone + listtwo
\end{lstlisting}

\begin{lstlisting}[caption={Combine multiple lists},label={lst:multiplelists}]
def merge(*lists):
    rslt = [""]
    for idx in range(len(lists[0])):

        r = []
        for s in rslt:
            for l in lists:
                r.append(s + l[idx])
        rslt = r
    return rslt
\end{lstlisting}
    
    \item Code with more details: The accepted answer invokes library APIs to implement the corresponding functionality, but the non-accepted code snippet implements the functionality itself. It has more implementation details, the developer turn to reuse it. For example, in the accepted answer, python's \code{sort} function is invoked to sort a list, while a quick sort algorithm is implemented to sort the list in the non-accept answer; and
    
    \item More efficient code: The non-accepted code snippet is modified on the accepted code snippet to make it more efficient. For example, in the accepted answer, the Fibonacci sequence is calculated using recursion as shown in the List. \ref{lst:fibrecur}. But in the unaccepted answer, a matrix is used for the calculation shown in the List. \ref{lst:fibmatrix};
    
\begin{lstlisting}[caption={Calculate Fibonacci sequence by recursion},label={lst:fibrecur}]
def fib(n):
if n <= 1:
    return n
else:
    return fib(n - 1) + fib(n - 2)
\end{lstlisting}
    
\begin{lstlisting}[caption={Calculate Fibonacci sequence using matrix},label={lst:fibmatrix}]
import numpy as np
def fib_matrix(n):
    Matrix = np.matrix([[0, 1], [1, 1]])
    vec = np.array([[0], [1]])
    return np.matmul(Matrix ** n, vec)
\end{lstlisting}
\end{itemize}

The average number of votes for Type 1 reused code snippets from the non-accept answer section is 11.25, while Type 2,3 is 5.33. From the number of votes for reused code snippets, we can observe that the higher the vote count, the more likely the code snippets are to be reused. It is worth noting that regardless of which part of the Stack Overflow post code snippets appear in, if it is reused by Type-1, then its average vote number is significantly greater than reused by Type-2,3. This phenomenon shows that code snippets with high votes are of higher quality and are more likely to be reused without modification. In contrast, code snippets with low votes are more likely to need to be modified before they are reused.

\begin{tcolorbox}[title=Answer to RQ2 (Which part reused?),boxrule=1pt,boxsep=1pt,left=2pt,right=2pt,top=2pt,bottom=2pt]

\noindent$\bullet$ For code reuse from the question section, we find that there are more reused code snippets from the question section than the answer section. We summarize three reasons by manually examining why these code snippets are being reused;

\noindent$\bullet$ For code reuse from the accepted answer section, we find that code snippets reused from the accepted answers represent 37.84\% of all answers. The average acceptance rate of all the answers in our data is 34.57\%. Similarly, we summarize three reasons why the accepted answer's code snippets are reused;

\noindent$\bullet$ For code reuse from the non-accepted answer section, we find that there are 1,148 (1.64\%) code snippets reused by Type-1 and 17,716 (25.28\%) reused by Type 2,3. There are four common reasons why developers reuse code from non-accept answer sections.

In summary, we find that when a post's vote is high, the code snippets are more likely to be Type-1 reused; low votes code snippets are more like to be reused by Type-2,3.

\end{tcolorbox}

\subsection{RQ3: Why do developers reuse code from Stack Overflow?}

\noindent\textbf{Motivation.} In RQ1, we find that developers are more prone to reuse code snippets from Stack Overflow. Following that research question, we need to explore the motivation of code reuse on Stack Overflow. By answering this RQ, we are able to provide insights into how developers benefit from reusing code snippets from Stack Overflow.

% On code reuse from StackOverflow: An exploratory study on Android apps

\noindent\textbf{Methodology.} To answer this RQ, we perform a qualitative analysis. Specifically, we ask two authors of this paper to manually inspect the motivations for reusing code snippets from Stack Overflow. To have a confidence level of 99\%, 1000 clone pairs were randomly selected from Type-1 and Type-2,3 clone pairs for qualitative analysis. One clone pair includes the related Stack Overflow post and the related Jupyter Notebook file. Second, we ask two authors of this paper to manually analyze the clone code pairs to obtain the following information, including context, description, and the functionality of the code snippets, which are used to categorize the code snippets. If discrepancies exist, another author of this paper is involved in the discussion until a consensus is reached. We use Cohen’s Kappa coefficient\cite{cohen1960coefficient} to measure the agreement of two raters.

\noindent\textbf{Results.}
After manual classification, we group the reuse reasons into six different categories. The result is shown in Table. \ref{tab:codereusereason}. We observe that Jupyter Notebook developers often reuse code snippets from Stack Overflow for different purposes. The two common reasons for reuse are adding new features and using APIs. That adds up to 615 and 530 for Type-1 and Type-2,3 code reuse, accounting for more than half of the reasons given. This phenomenon is consistent with our intuition that most developers reuse code snippets from Stack overflow when they have trouble implementing a feature or encountering an unfamiliar API. It is important to note that the import libraries is also part of the reuse reason and cannot be ignored. The most commonly reused code snippets are shown in List \ref{lst:importlib}, which shows numerical calculation, data processing, and the graphing module is the most frequently used by Jupyter Notebook developers in our data collection.
\begin{lstlisting}[caption={Most reused snippets of imported libraries},label={lst:importlib}]
    import pandas as pd
    import numpy as np
    import matplotlib.pyplot as plt
    import seaborn as sns
    %matplotlib inline
\end{lstlisting}
When comparing the difference between code reuse causes of Type-1 and Type-2,3, we can observe that the cause of module import in Type-1 is 289. The number of Type-2,3 is 1.2 times higher than Type-1, with 334. This is because, unlike Type-2,3 code reuse, Type-1 has strict requirements on the order of statements, and Type-2,3 code reuse is more concerned with semantic similarity.  Since the order of importing modules in code snippets does not affect the semantics, there are more causes in Type-2,3 code reuse caused by importing modules. Among the reasons for reusing the initialization data, Type-2,3 code reuse is twice as high as Type-1 code. One possible explanation is that when developers process data, although the data itself is the same, the way of processing is different. Therefore, there is a lot of code considered to reuse the initialization data for Type 2,3 reuse. We use Cohen’s Kappa coefficient \cite{cohen1960coefficient} to measure the agreement between two raters. The result is +0.85, showing good agreement between the two raters.

\begin{tcolorbox}[title=Answer to RQ3 (Reasons for Code Reuse),boxrule=1pt,boxsep=1pt,left=2pt,right=2pt,top=2pt,bottom=2pt]
Jupyter Notebook developers reuse StackOverflow code for the following reasons: (1) adding new features; (2) using the APIs; (3) importing libraries; (4) reusing the initialization data; and (5) reusing test cases. The most common reason for reusing StackOverflow code snippets is to add new features.
\end{tcolorbox}

\begin{center}
\begin{table*}[b]%
\caption{Reason for code reuse.\label{tab:codereusereason}}
\centering
\begin{tabular*}{300pt}{@{\extracolsep\fill}lcc@{\extracolsep\fill}}
\toprule
&\multicolumn{2}{@{}c@{}}{\textbf{Frequency}}  \\\cmidrule{2-3}
\textbf{Category} & \textbf{Type-1}  & \textbf{Type-2,3}     \\
\midrule
Adding new features & 368          & 223            \\
API usage           & 247          & 307            \\
Importing libraries    & 289          & 334           \\
Data initialization    & 48           & 98             \\
Testing             & 29           & 26             \\
Other               & 19           & 12             \\
\bottomrule
\end{tabular*}
\end{table*}
\end{center}

\subsection{RQ4: What is the quality of code snippets provided in answers on Stack Overflow?}

\noindent\textbf{Motivation.} According to the results in RQ1, we find that existing Jupyter Notebook developers reuse code snippets from Stack Overflow. Furthermore, existing work \cite{meldrum2020understanding} finds that there exist low-quality code snippets on Stack Overflow. Thus, reusing code snippets from Stack Overflow can introduce security risks to the Jupyter Notebook. In this RQ, we intend to evaluate the quality of the code snippets provided in answers on Stack Overflow.

\noindent\textbf{Methodology.} To answer this RQ, we evaluate the quality of reused code snippets in Jupyter Notebook from four dimensions: (1) reliability and conformance to programming rules, (2) readability, (3) performance, and (4) security. The details of these four dimensions are presented in Table. \ref{tab:quality-evaluation}.

\newcolumntype{C}[1]{>{\centering\arraybackslash}p{#1}}
\begin{table*}[!htbp]
\caption{Description of the code snippet's quality  four dimensions}
\centering
\begin{tabularx}{0.8\textwidth}{| c | X|}
\hline
\textbf{Quality Dimension} & \textbf{\centering Description}  \\ \hline
\textbf{Reliability and Conformance} &  Code snippets should be able to compile and contain no bugs and errors. Furthermore, the code snippets should also conform to accepted programming rules.
\\ \hline
\textbf{Readability}  &  Code snippets should follow standard Python readability conventions to ensure they can be easily understood and maintained. 
\\ \hline
\textbf{Performance} &   Performance and efficiency should be considered when reusing code snippets. For example, has the code snippet offered in an answer improve the performance (e.g., saving processing steps)
\\ \hline
\textbf{Security}  &  Reusing code snippets should consider the security issues of the snippets.
\\ \hline
\end{tabularx}

\label{tab:quality-evaluation}
\end{table*}

\begin{table*}[!htbp]
\caption{Supported Performance related Violation type in PyLint}
\centering
\begin{tabularx}{0.8\textwidth}{|c|X|}
\hline
\textbf{Violation Type} & \textbf{Description}  \\ \hline

self-assigning-variable  & A statement in the code that assigns a variable's value to itself. For example, a statement like ``foo = foo''.
\\ \hline % W0127

comparison-with-itself  &  A statement that compares a variable with itself, for example, ``if foo $>$ foo: ....''
\\ \hline % R0124

simplifiable-condition  &  A conditional statements that can be simplified, for example ``if 2 $>$ 1 and a $>$ b: ... '', can be simplified to ``if a $>$ b: ....''.
\\ \hline % R1726

condition-evals-to-constant  &  A conditional statements that can be simplified to constants, a special case of R1726. For example ``if 2 $>$ 1:...''.
\\ \hline % R1727

consider-using-in  &  Checking if a variable is equal to one of many values, it is better to combine those values into a tuple or set and use the ``in'' keyword instead of comparing variables to values one by one. For example, you should use ``a in (1, 2)'' instead of ``a == 1 or a == 2''.
\\ \hline % R1714

consider-merging-isinstance  &  When using the \code{isinstance} function consecutively, it can be merged into one. For example, ``isinstance(value, int) or isinstance(value, float)'' can be merged to ``isinstance(value, (int, float))''.
\\ \hline % R1701

consider-using-generator  &  If a container is large, using a generator will bring better performance. For example ``list([0 for y in list(range(10))])'' can be refactored to ``list(0 for y in list(range(10))) \# using generator''
\\ \hline % R1728

use-a-generator  &  It is more efficient to use a generator instead of comprehension when invoke ``any'', ``all'', ``max'', ``min'', ``sum'' functions. For example, for ``all([randint(-5, 5) $>$ 0 for \_ in range(10)])'' statement, it is better to use generator like ``all(randint(-5, 5) $>$ 0 for \_ in range(10))'', because it can cut the execution tree and exit directly at the first element that is \code{False}.
\\ \hline % R1279

consider-using-join  &  When concatenating strings, ``str.join(sequence)'' should be used instead of using for-loop iteration. For example,  ``''.join(["a", "b"])'' is more efficient than ``s = '' for c in ["a", "b"]: s += a''.
\\ \hline % R1713

unnecessary-dict-index-lookup  &  When enumeration is performed on the dict, the value and its index can be directly obtained, and there is no need to use the index to obtain the value. For example, you should use ``d = \{'a': 1, 'b': 2\} for key, value in enumerate(d):    print(value)'' rather than ``d = \{'a': 1, 'b': 2\} for key, value in enumerate(d):    print(d[key])''.
\\ \hline % R1733

unnecessary-list-index-lookup  & When enumeration is performed on the list, the value and its index can be directly obtained, and there is no need to use the index to obtain the value. For example, ``l = ['a', 'b'] for index, c in enumerate(l):    print(c)'' is better than ``l = ['a', 'b'] for index, c in enumerate(l): print(letters[index])''.
\\ \hline % R1736
use-sequence-for-iteration  &  When iterating over values, sequence types (e.g., lists, tuples, ranges) are more efficient than sets. For example, ``l = [1, 2] for i in enumerate(l): print(i)'' is efficient than ``l = \{1, 2\} for i in enumerate(l): print(i)''
\\ \hline % C0208
use-list-literal  &  When creating a new list, it is faster to use [] instead of list(). Because it avoids an additional function call
\\ \hline % R1734
consider-using-tuple & Consider using an in-place tuple() instead of list(). Due to optimizations by CPython, there is no performance benefit from it.
\\ \hline % R6102
\end{tabularx}

\label{tab:security-checker}
\end{table*}

Specifically, evaluating the four dimensions for the quality of the code snippets. For reliability and conformance, we leverage \textit{PyLint} \cite{pylint} for the task. Specifically, PyLint checks errors in the code and looks for code smells in the code. For readability, we leverage \textit{pycodestyle} \cite{pycodestyle} to evaluate the code snippets reused from Stack Overflow. \textit{Pycodestyle} checks Python code against the style conventions in PEP 8 \cite{pep08}, which is the style guide for Python code. For performance, we leverage \textit{PyLint} \cite{pylint} to check potential performance-related issues. In \textit{PyLint}, it supports different types of checkers to find potential defects in the code snippets. We select all performance related checkers in PyLint and leverage them to detect potential defects in code snippets. PyLint supports 50 checkers, such as Basic checker, Refactoring checker, and Type checker. For each check, it checks several possible defects in the code. We select all related defects from all 50 checkers. As shown in Table. \ref{tab:security-checker}, in summary, we obtained 15 issues from these checkers related to performance.

For security, we leverage the \textit{bandit} tool \cite{bandit} to find common security issues in reused Python snippets. 

\noindent\textbf{Results.}
We scan all Jupyter Notebook code snippets reused from Stack Overflow posts and used the tools noted in the methodology. After scanning 193,248 code snippets, we find 1,528,844 violations. That means, on average, a code snippet has 7.91 code quality violations. The general distribution of violations across the four quality attributes is summarized in Table \ref{tab:violationssummary}. Note that we have removed the violations that are unrelated to Jupyter Notebook code snippets. For example, the code snippet shown in the List \ref{lst:violationexample}, after scanning with pylint, this code snippet is reported as having a pointless-statement violation. However, the ``df\_a`` variable defined in the code snippet is used later, and we argue that this code snippet is not pointless-statement and removed from the violations. Next, we discuss the violation of four quality attributes in more detail.

\begin{lstlisting}[caption={Example of violations not related to Jupyter Notebook code snippet},label={lst:violationexample}]
raw_data = {
        'first_name': ['Alex', 'Amy', 'Allen', 'Alice', 'Ayoung'], 
        'last_name': ['Anderson', 'Ackerman', 'Ali', 'Aoni', 'Atiches']}
df_a = pd.DataFrame(raw_data, columns = ['first_name', 'last_name'])
df_a
\end{lstlisting}

\begin{center}
\begin{table}[t]%
\centering
\caption{Summary of violations for the four quality dimensions.\label{tab:violationssummary}}%
\begin{tabular*}{500pt}{@{\extracolsep\fill}lcccc@{\extracolsep\fill}}
\toprule
\textbf{Number of violations} & \textbf{Reliability and conformance}  & \textbf{Readability}  & \textbf{Performance}  & \textbf{Security} \\
\midrule
Median               & 2                           & 4                           & 0           & 0                            \\
Average              & 2.63                    & 5.23                    & 0.02        & 0.04                        \\
Maximum              & 481                         & 1,097                              & 11          & 17                           \\
Total                & 507,298              & 1,010,342                      & 4,165       & 7,039                        \\
\bottomrule
\end{tabular*}
\end{table}
\end{center}

\subsubsection{Reliability and Conformance}
There are 507,298 violations relate to reliability and conformance, accounting for 33.18\% of all violations. Of the code snippets analyzed, 144,486 (78.22\%) have violations. The average number of violations per snippet is 2.74. The maximum number of violations per snippet is 481, indicating the prevalence of reliability and conformance violations in the reused snippets.
    
Furthermore, we explore the distribution of the number of violations in code snippets. 123,079 (85.18\%) snippets contain 1-5 violations, 11,788 (8.16\%) snippets contain 6-10 violations, 4,223 snippets (2.92\%) have 11-15 violations, 2267 snippets (1.57\%) have 16-20 violations. The number of code snippets with more than 20 violations is 3129 (2.17\%). 

The top ten reasons for the most common code snippet violation are shown in the Tab \ref{tab:mostpylint}. The \code{line-too-long} is the most common reason for violations, with 113,364 (22.35\%). PEP 8 suggests for flowing long blocks of text with fewer structural restrictions (docstrings or comments), the line length should be limited to 72 characters. \cite{style-guide}. These ten violations can be divided into 5 categories: violations related to code style (trailing-whitespace, bad-indentation, line-too-long, redefine-outer-name), violations against best practices (consider-using- f-string), wrong import violations (wrong-import-position, wrong-import-order), missing documentation violations (missing-function-docstring, missing-class-docstring) and syntax errors (syntax-error). Violations related to code style are the category containing the most violations, with 321,047 representing 63.28\% of all violations.

\begin{center}
\begin{table}[t]%
\centering
\caption{Most Common Reliability and Conformance Violations.\label{tab:mostpylint}}%
\begin{tabular*}{230pt}{@{\extracolsep\fill}lc}
\toprule
\textbf{Violation Type} & \textbf{Number of Violations}  \\
\midrule
line-too-long              & 113,364                        \\
trailing-whitespace        & 98,522                         \\
bad-indentation            & 97,383                         \\
missing-function-docstring & 51,681                         \\
syntax-error               & 50,297                         \\
consider-using-f-string    & 25,921                         \\
wrong-import-order         & 14,292                        \\
redefined-outer-name       & 11,778                         \\
missing-class-docstring    & 7,906                          \\
wrong-import-position      & 5,740                          \\
\bottomrule
\end{tabular*}
\end{table}
\end{center}

\subsubsection{Readability}

Readability-related violations is the highest of the four quality dimensions, with 1,010,342 (66.09\%). Of the code snippets analyzed, 129,075 (69.87\%) have violations, with an average of 5.23 violations per snippet, and the snippet with the most readability violations had 1,097 violations. Compared with reliability and conformance, there are more readability violations but fewer code snippets involved.

Similarly, the number of violations among code snippets is calculated; 84,116 (65.17\%) code snippets have 1 to 5 violations, 22,424 (17.37\%) code snippets have 6 to 10 violations, and 7704 (5.97\%) code snippets have 11 to 15 violations. There are 3930 (3.04\%) code snippets with 16 to 20 violations and 10901 (8.44\%) code snippets with more than 20 violations.

To figure out why code snippets violate readability, we compile the top ten most common reasons for violations, as shown in the Table. \ref{tab:mostpycodestyle}. We observe that missing-whitespace is the most common cause of style violation, with a total of 399,767 (39.57\%). Among the top ten causes of violations, Whitespace caused the most violations (missing-whitespace, trailing-whitespace, unexpected-spaces, blank-line-contains-whitespace, missing- operators-whitespace, missing-comment-whitespace) with 783,486 (77.54\%); Line length related violations (line-too-long) with 113,364 (11.22\%); Indentation caused violations are indentation-not-multiple-of-four and under-indented, which add up to 128,201 (12.69\%); violations related to import (module-import-should-be-top) have a 50987(5.04\%).

\begin{center}
\begin{table}[t]%
\centering
\caption{Most Common Readability Violations.\label{tab:mostpycodestyle}}%
\begin{tabular*}{250pt}{@{\extracolsep\fill}lc}
\toprule
\textbf{Violation Type} & \textbf{Number of Violations}  \\
\midrule
missing-whitespace                                                            & 399,767                        \\
line-too-long                                                                 & 113,364                        \\
trailing-whitespace                                                           & 98,522                         \\
indentation-not-multiple-of-four                                              & 90,573                         \\
unexpected-spaces                                                             & 85,894                         \\
blank-line-contains-whitespace & 83,157                         \\
operators-missing-whitespace   & 79,315                         \\
module-import-should-be-top                                                         & 50,987                         \\
under-indented                & 37,628                         \\
comment-missing-whitespace    & 36,831                         \\
\bottomrule
\end{tabular*}
\end{table}
\end{center}

\subsubsection{Performance}
Pylint finds 4,687 performance violations, with an average of one violation per 0.02 code snippets. Performance violation is the least of our four quality dimensions, showing that Jupyter Notebook developers care more about the performance of reused code than other quality dimensions, such as code style. Additionally, there are 2,182 (0.14\%) code snippets with performance violations, the largest number of violations being 11. There are 2050 (93.95\%) code snippets with a violation count between 1 and 5, 130 (5.96\%) code snippets with a violation count between 6 and 10, and only 2 (0.09\%) code snippets with a violation count greater than 10.

We count the number of all causes of performance violations, which are shown in Table. \ref{tab:performance}. The most performance violations we observed (1,621 or 34.59\%) are due to consider-using-tuple violations, with self-assigning-variable in second place with 1357 (28.95\%) and use-list-literal with 811 (17.3\%). The rest of the violations are below 10\%. We recommend that Jupyter Notebook developers make more efficient use of built-in \textit{list} types when reusing code snippets and avoid assigning a variable's value to itself.

\begin{center}
\begin{table}[t]%
\centering
\caption{Number of performance violations.\label{tab:performance}}%
\begin{tabular*}{250pt}{@{\extracolsep\fill}lc}
\toprule
\textbf{Violation Type} & \textbf{Number of Violations}  \\
\midrule
consider-using-tuple          & 1,621                          \\
self-assigning-variable       & 1,357                          \\
use-list-literal              & 811                           \\
consider-using-generator      & 393                           \\
consider-using-in             & 381                           \\
unnecessary-list-index-lookup & 56                            \\
consider-merging-isinstance   & 34                            \\
comparison-with-itself        & 14                            \\
condition-evals-to-constant   & 12                            \\
simplifiable-condition        & 2                             \\
consider-using-join           & 2                             \\
unnecessary-dict-index-lookup & 2                             \\
use-sequence-for-iteration    & 2                             \\
use-a-generator               & 0                             \\
\bottomrule
\end{tabular*}
\end{table}
\end{center}

\subsubsection{Security}
We find 7,039 (0.46\%) security violations using bandit, an average of one security violation for every 0.04 code snippets. The remaining 27 snippets (0.65\%) had more than 5 violations. 

We count the top ten most common reasons for security violations, which are shown in the Table. \ref{tab:banditviolation}. Unlike the previous three quality categories, the bandit has graded each violation by severity, and it can be observed that among the most common reasons for violations, 5171 have a low severity, accounting for 73.46\% of all violations, while the medium has 1254 or 17.82\%. Violations with high severity are not among the top ten common violations, with 66 accounting for 0.94\% of all security violations. These data suggest that the problem of harmful reuse of code in Jupyter Notebook is not severe. The top ten most common reasons for violation can be divided into 3 categories, invokes blacklists function(use-standard-pseudo-random-generators, unexpected-url-open-parameter, use-insecure-function), import blacklists module(use-pickle -module, use-subprocess-module, use-xml-parse-untrusted-data) and misc violation (assert-used, insecure-use-temp-file, try-except-pass-found, possible-hardcoded-password). The highest number of misc violations is 2740, invokes blacklists function is 2036, and the lowest is the import blacklists module is 1649.

\begin{center}
\begin{table}[t]%
\centering
\caption{Most Common Security Violations.\label{tab:banditviolation}}%
\begin{tabular*}{350pt}{@{\extracolsep\fill}lcc}
\toprule
\textbf{Violation Type} & \textbf{Number of Violations} & \textbf{Severity} \\
\midrule
assert-used                           & 1705                          & low               \\
use-pickle-module                     & 1464                             & low               \\
use-standard-pseudo-random-generators & 1130                            & low               \\
unexpected-url-open-parameter         & 545                            & medium            \\
use-insecure-function                 & 361                           & medium            \\
insecure-use-temp-file                & 348                            & medium            \\
possible-hardcoded-password           & 347                           & low               \\
try-except-pass-found                 & 340                             & low               \\
use-subprocess-module                 & 103                             & low               \\
use-xml-parse-untrusted-data          & 82                             & low               \\ 
\bottomrule
\end{tabular*}
\end{table}
\end{center}

\subsubsection{Type-1 and Type-2,3 reused code snippets' quality}
The result for the number of violations for different code reuse in four quality dimensions is shown in Table. \ref{tab:typeviolation}. We find that Type-1 reuse code snippets have 38,908 reliability violations, with an average of 1.68 violations per code snippet. 

Type-2,3 reuse code snippets have 490,629 reliability violations. There are 2.65 reliability violations per code snippet. On average, each Type-2,3 reuse code snippet has 33.84\% more reliability violations than Type-1 reuse code snippet. In the readability quality dimension, there are 63,608 violations for Type-1 code reuse snippets and 978,559 violations for Type-2,3 code reuse snippets. 

On average, Type-2,3 reused code snippets contain 2.06 more readability violations than Type-1 reused code snippets. For the number of average performance violations, Type-2,3 reused code snippets are 0.01 larger than that in Type-1 code snippets. The Type-1 and Type-2,3 reused code snippet's average violations are the same on the security dimension. We observe that Type-2,3 reused code snippets are more likely to have reliability and readability violations. On the performance and security dimensions, there is no significant difference in the number of average violations per code snippet for different types of code clones.

\begin{center}
\begin{table}[t]%
\centering
\caption{The number of code violations in different reuse Types.\label{tab:typeviolation}}%
\begin{tabular*}{500pt}{@{\extracolsep\fill}lcccc@{\extracolsep\fill}}
\toprule
\textbf{Quality Dimension} & \textbf{Type-1 Violations}  & \textbf{Violations Per Snippet}  & \textbf{Type-2,3 Violations}  & \textbf{Violations Per Snippet} \\
\midrule
Reliability & 38,908             & 1.98                   & 490,629              & 2.65                   \\
Readability & 63,608             & 3.24                   & 978,559              & 5.30                   \\
Performance & 237               & 0.01                   & 4,034                & 0.02                   \\
Security    & 745               & 0.04                   & 6,662                & 0.04                          \\
\bottomrule
\end{tabular*}
\end{table}
\end{center}

\begin{tcolorbox}[title=Answer to RQ4 (Code Quality),boxrule=1pt,boxsep=1pt,left=2pt,right=2pt,top=2pt,bottom=2pt]
In this RQ, we measure the quality of the reused code in Jupyter Notebook from four dimensions. On average, a code snippet has 7.91 code quality violations. We find that reused code snippets have the highest number of violations on readability with an average of 3.39 violations per code snippet. In the reliability and conformance quality dimension, reused code snippets have the second-highest number of violations with an average of 2.74 violations per code snippet. There are only 0.04 Security violations and 0.02 Performance violations per code snippet. We find that Type-2,3 reused code snippets are more likely to have reliability and readability violations.
\end{tcolorbox}

\subsection{RQ5: Who reuses code from Stack Overflow?}

\noindent\textbf{Motivation.} Previous research \cite{Ohad:2014} demonstrated that reusing code can have negative effects on building software. This always associates with less experienced developers. For example, developers with less experience may adopt the Type-1 clone. Here, we intend to examine who reuses Stack Overflow code among developers. Furthermore, we also aim to investigate how experienced developers cope with code reuse.

\noindent\textbf{Methodology.} To answer this RQ, we adopt the same methodology presented in \cite{abdalkareem2017code}. Specifically, we first measure the experience of developers. Similar to previous works \cite{rahman2019snakes,tahir2018can}, we measure the number of commits from the start of the project to the time of code reuse. This helps us evaluate the experience of developers. One developer commits more often than another developer, he/she is more familiar with the project and is more experience. We normalize the experience of a developer as a percentage of the total number of commits he/she made to the project.

\begin{figure}[!htpb]
    \centering
    \includegraphics[width=0.5\textwidth]{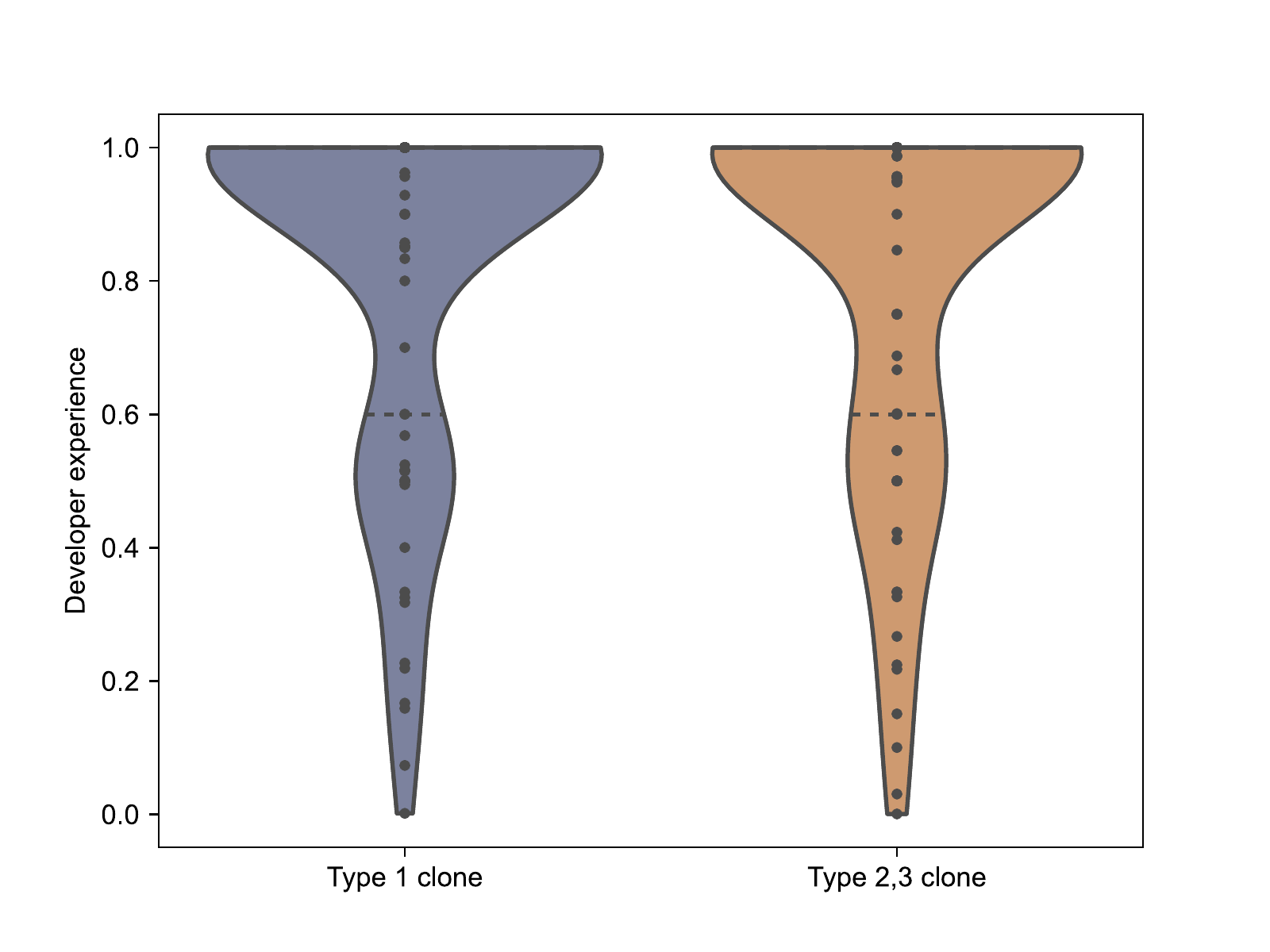}
    \caption{Distribution of the developers’ experience for developers who reuse code from StackOverflow}
    \label{fig:devexperience}
\end{figure}

\noindent\textbf{Results.}
We present the distribution of experience in projects by authors who reuse Stack Overflow code snippets, as shown in Figure \ref{fig:devexperience}. We observe the majority of developers who perform code reuse have an experience of 1 (81.72\% and 77.17\% for Type-1 and Type-2,3 code reuse, respectively), while there is also a partial distribution of data in the 0.4 to 0.6 range (4.25\% and 4.48\% for Type-1 and Type-2,3 code reuse respectively), and the developer's experience doesn't correlate significantly with the way of reuse. By inspecting the total number of developers in Jupyter Notebook projects where code reuse exists, we find the following possible reason for the distribution of developer experience with code reuse. First, many Jupyter Notebook projects have only one developer, which results in a large number of reusers with experience of 1. Second, in projects with more than one developer, there are also more projects with only two developers, so the developer experience tends to be distributed around 0.5.

For more detail, We obtain 149,541 Jupyter Notebook code snippets with developers' experience of 1. The number of one-author code snippets accounted for 98.14\% of them. We also count 43,707 code snippets with developer experience of less than 1, and projects with two developers accounted for 19.85\% of them. Meanwhile, there are 5,850 code snippets in which the developer is two, and the developer experience is in the 0.4 to 0.6 range. This represents 13.38\% of the clone pairs with developer experience of less than 1 and 67.41\% of the code snippets with developer experience in the range of 0.4 to 0.6.

\begin{figure}[!htpb]
    \centering
    \includegraphics[width=0.5\textwidth]{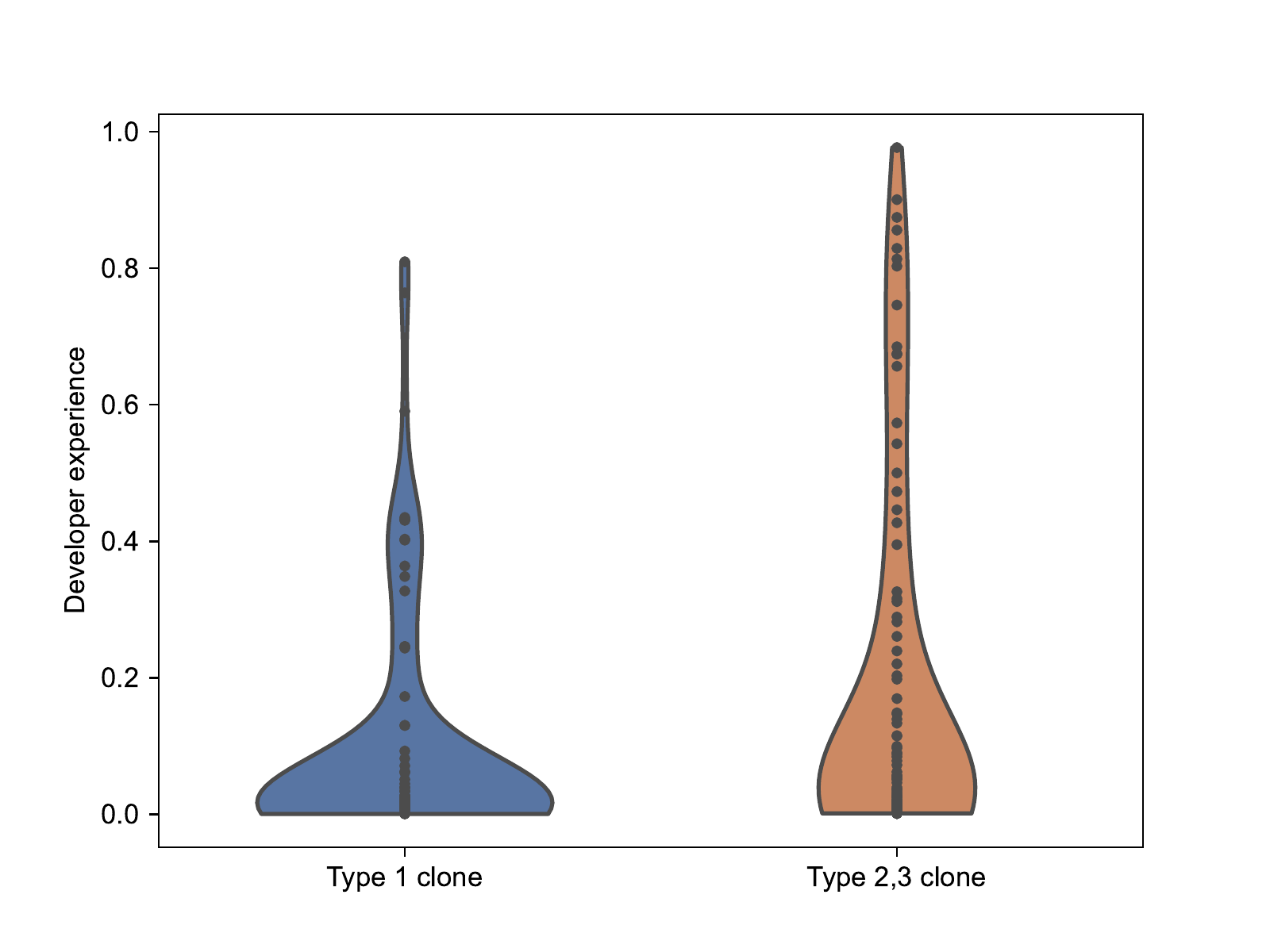}
    \caption{Distribution of the developers’ experience on teamwork project}
    \label{fig:devexperiencefilter}
\end{figure}

To focus on the distribution of developer experience in medium and large projects, we filter out small Jupyter Notebook projects by the total number of developers greater than 10. The results are shown in Fig. \ref{fig:devexperiencefilter}. We find that less experienced developers (less than 0.1 experience) are more likely to reuse code with the Type-1 clone, accounting for 83.28\% of all Type-1 code clones. The proportion of developers with Type 2,3 code reuse experience below 0.1 is 52.46\%. This result suggests that developers with less development experience on medium or large projects tend to resort to Type-1 code cloning for code reuse. Experienced developers turn to choose Type-2,3 code clone.

\begin{tcolorbox}[title=Answer to RQ5 (Who Reuse Code?),boxrule=1pt,boxsep=1pt,left=2pt,right=2pt,top=2pt,bottom=2pt]
When it comes to code reuse, most Jupyter Notebook developers have an experience of 1, and there is no significant difference in the reuse method because most of the Jupyter Notebook projects we collected are small personal projects. Furthermore, we find that the less experienced developers intend to reuse code using Type-1 code clone. In contrast, more experienced developers intend to reuse code with Type-2,3 code clones.

\end{tcolorbox}

\section{Lessons Learnt and Threats to Validity}\label{sec:lessons}
In this section, we summarize our findings from this study and provide suggestions for developers. Then we present the threats to the validity of our work.
\subsection{Lessons Learnt}
We find that 59,942 of the 188,302 (31.83\%) Github repositories we collected contain code reuse. This shows that code reuse is widespread in Jupyter Notebook. To make it easier for Jupyter Notebook programmers to reuse code on Stack Overflow, we have summarised our findings in the following three aspects.

\noindent\textbf{(1) Reasons for reuse code} Based on our manual analysis, we find that adding new features, importing modules ,and using APIs are the three most common reasons for reuse. This suggests that developers turn to Stack Overflow for help when they are faced with a need for knowledge they are not familiar with. We also find that developers do not always reuse code snippets from the accepted answer, as the question section or unaccepted answers may have more efficient implementations or code that is easier to reuse. So we recommend that when looking for code to reuse on Stack Overflow, developers should not focus too much on the accepted answer, as there may be more suitable code in the question or unaccepted answer section.

\noindent\textbf{(2) Reuse code quality} Our experiments on code quality show that for Jupyter Notebook code snippets in reused pairs. On average, there are 7.91 code quality violations per code snippet. This indicates that code quality violations are common in Jupyter Notebook reuse code snippets. After reusing a Stack Overflow code snippet, we recommend that developers use code quality checkers to check the code. Avoid the situation where reusing external code degrades the quality of the program.

\noindent\textbf{(3) Code reuser's experience:} We find that more experienced developers tend adopt Type-2,3 code clone in medium and large projects. Our experiments have shown that Type-2,3 code snippets are more likely to have reliability and readability violations. Thus, for large-size projects, developers should carefully inspect code reliability and readability violations for reused type 2,3 code snippets.

\subsection{Threats to Validity}
\textbf{Threats to internal validity.} We use SourcererCC to detect Type-2,3 code clones. However, the accuracy of the SourcererCC detection limits our identification of Type-2,3 clone pairs. In some cases, SourcerCC detected code snippet pairs that did not have Type-2,3 clones as code clones. Furthermore, SourcerCC may miss code clone pairs for Type-2,3. To mitigate this problem, we manually sampled several code snippet pairs reported by SourcerCC and code snippets not reported as clones by SourcerCC. In all cases, the reported clones were actual clones, and there was no underreporting.

When looking at why developers reuse Stack Overflow, we use manual classification as this task is hard or even impossible to automate. To avoid errors associated with manual classification, we ask two authors to classify independently and then discuss the results to reach a consensus on the classification. We use Cohen's Kappa to calculate an inter-rater agreement of 0.85, which indicates a high degree of agreement between those two.

\noindent\textbf{Threats to external validity.} There is a possibility that the results of this study are not generalize. 

\noindent$\bullet$ First, we focus on reusing code snippets from Stack Overflow, which is only one of many Q\&A websites, so there is a possibility that the results cannot be generalized to all Q\&A websites. 

\noindent$\bullet$ Second, although Jupyter Notebook Note is most commonly used as a tool for analyzing data\cite{koenzen2020code,pimentel2019large}, there is a possibility that our findings may not be generalizable to other applications (e.g. Polynote or R Notebook).
\section{Related Work}\label{sec:relatedwork}
We discuss the related work from three aspects: studies on StackOverflow, studies on Jupyter Notebook, and studies related to code reuse and clone detection.

\noindent\textbf{Works related to StackOverflow.} An et al. \cite{an2017stack} studied whether app developers respect license terms when reusing code from StackOverflow posts by inspecting 399 Android apps. Uddin et al. \cite{uddin2020mining} proposed a framework named Opiner to mine API usage patterns from StackOverflow posts, which can be used to assist developers in programming tasks and reusing code segments from StackOverflow posts. Ponzanelli et al. \cite{ponzanelli2013seahawk} developed an Eclipse plugin to generate queries for StackOverflow according to the programming context in the IDE. Some works care about the code quality of code segments on StackOverflow posts. Specifically, Wu et al. \cite{wu2019developers} discussed how developers utilize code segments from StackOverflow. They conduct an exploratory study on 289 files from 182 open-source projects, which contain source code that has an explicit reference to a StackOverflow post. They found the top 3 barriers that make it difficult for developers to reuse code from StackOverflow. Ragkhitwetsagul et al. \cite{ragkhitwetsagul2019toxic} conducted a large-scale empirical study for online code clones between Stack Overflow and 111 Java open source projects to understand: (1) code clone practices; (2) code clone patterns; (3) out-dated code clones; and (4) software licensing violations. Meldrum et al. \cite{meldrum2020understanding} conducted a large-scale study on code snippet quality for code snippets from StackOverflow. Rahman et al. \cite{rahman2019snakes} discussed the insecure code practices for Python code on StackOverflow. Other works focus on the code smell detection for code segments on StackOverflow \cite{tahir2018can, shcherban2020automatic,duijn2015quality}. 

\noindent\textbf{Works related to Jupyter Notebook} Pimentel et al. conducted a large-scale study on the quality and reproducibility of Jupyter notebooks \cite{pimentel2019large}. They studied 1.4 million notebooks from GitHub, showing that only 24.11\% of Jupyter notebooks can be executed without exception, and only around 4\% of notebooks can be reproduced with the same results. Koenzen et al. \cite{koenzen2020code} explored the way of code duplications in Jupyter notebooks and identified the potential barriers to code reuse. Besides, Wang et al. \cite{Wang:2020ASE} first studied whether existing notebooks can be executed successfully (i.e., reproducibility). Then, they proposed a prototype named Osiris, which takes a notebook as an input and outputs the possible execution schemes to reproduce the notebook. Wang et al. \cite{Wang:2021} developed SnifferDog to restore the execution environments for executing Jupyter notebooks. Specifically, SnifferDog first collects the APIs of Python packages to build the database and then analyzes the notebooks to determine the candidate packages and versions. Wang et al. 's work \cite{wang2020better} conducted a preliminary study on code quality for Jupyter notebooks. They found that the existing notes are with poor quality codes, which requires quality control on Jupyter notebooks. Other work is developing tools to improve the readability or efficiency of Jupyter Notebook\cite{9825898,10.1145/3510003.3510144}.

\noindent\textbf{Works related to clone detection and reuse.}
Roy et al. \cite{roy2008nicad} developed NiCAD, which leverages a text-based approach to detect Type-1 and Type-2 code clones. Saini et al. \cite{saini2018oreo} developed a code clone tool Oreo by leveraging machine learning, information retrieval, and software metrics. Lopes et al. \cite{lopes2017dejavu} studied code clones on a corpus of 4.5 million non-fork projects on GitHub. These projects represent over 428 million lines of code in Java, C++, Python, and JavaScript. Code reuse has been widely studied \cite{gharehyazie2019cross, yang2017stack,abdalkareem2017code, ahmad2019impact,haefliger2008code,9425939,feitosa2020code,yang2016query,7180116,235481,diamantopoulos2019towards}. Specifically, Ossher et al. \cite{ossher2011file} developed a line-level code clone detection tool, which is used upon the Sourcerer Repository. Among over 13,000 projects in Sourcerer Repository, they found that over 10\% of files are cloned. Gharehyazie et al. \cite{gharehyazie2019cross} detected cross-project clones among 8,599 projects on GitHub for Type-1 and 2 clone. Yang et al. \cite{yang2017stack} presented a large-scale study on 909k non-fork Python projects on GitHub, and 1.9 million Python snippets captured in Stack Overflow to learn the code clone on these projects. Abdalkareem et al. \cite{abdalkareem2017code} studied the code reuse practice by studying how developers reuse code snippets from Stack Overflow when building Android apps. Their study provides the potential impact of code reuse from StackOverflow on building apps. Ahmad et al.\cite{ahmad2019impact} measured the impact of the code snippets from Stack Overflow in  GitHub projects during the evolution, including prior to the addition of the snippet, immediately after the addition of the snippet and a longer time after the addition of the snippet. They found that almost 70\% of the cases where the copied snippet affected the cohesion of the project.

\section{Conclusion} \label{sec:conclusion}
%In this paper, we do an exploratory study of Jupyter Notebook's reuse of Stack Overflow code snippets. Our first study looked at how much code reuse happened in Jupyter Notebooks. 3.26\% of the code snippets in the data we collected were related to code clones. Furthermore, we analysed the reason for code reuse in terms of Stack Overflow attributes, developer experience, and developer intent. We found that (1) code snippets with higher votes are more likely to be reused by Type 1 code, while code snippets with lower votes are more likely to be reused by Type 2 and 3 code; (2) for small projects, most developers have an experience of 1, for large projects, less experienced developers use Type 1 code reuse, while experienced developers use Type 2,3 code reuse; (3) we found that the three most common reasons developers reuse code snippets were adding new features, importing modules, and API usage. Finally, we looked at the quality of reused code snippets. We found a total of 1,528,844 violations, with an average of 7.91 violations per code snippet; Readability violations were the most numerous, accounting for 66.09\% of the total violations.

%Based on the results of our exploratory research, we recommend that Stack Overflow conducts more rigorous code reviews of code snippets that are more likely to be reused to improve code quality and that developers be aware of the impact of reusing snippets on code quality when reusing off-the-shelf code.

In this paper, we investigate the reuse of code from Stack Overflow in Jupyter Notebook in Stack Overflow. We find that 3.26\% of code snippets are related to code reuse. Then, by analyzing the source of cloned code snippets, we find that code snippets come from the question more than the answer section. In addition, we summarise the three most common reasons for code reuse: adding new features, using APIs, and importing modules. By examining the quality of the reused code snippets, we find an average of 7.91 violations per snippet, with readability violations being the most frequent, accounting for 66.09\% of the total violations. Finally, we learn that in medium to large projects, less experienced developers are more likely to adopt Type 1 code clones than more experienced developers. Our research provides insight into code reuse on Stack Overflow in Jupyter Notebook.

\nocite{*}% Show all bib entries - both cited and uncited; comment this line to view only cited bib entries;
\bibliography{wileyNJD-AMA}%

\clearpage

\end{document}